\documentclass[10pt,twocolumn]{IEEEtran}
\usepackage{soul}
\usepackage{graphicx}
\usepackage{amsmath,amssymb}
\usepackage{amsfonts}
\usepackage{algorithm}
\usepackage{algorithmic}
\usepackage{array}
\newcommand{\PreserveBackslash}[1]{\let\temp=\\#1\let\\=\temp}
\newcolumntype{C}[1]{>{\PreserveBackslash\centering}p{#1}}
\newcolumntype{R}[1]{>{\PreserveBackslash\raggedleft}p{#1}}
\newcolumntype{L}[1]{>{\PreserveBackslash\raggedright}p{#1}}
\usepackage[usenames]{color}
\usepackage{colortbl,booktabs}
\usepackage{tabularx}
\usepackage{multicol}
\usepackage{multirow}
\usepackage{multicol}
\usepackage{booktabs}
\usepackage[Symbol]{upgreek}
\usepackage{subfigure}
\usepackage{bm}
\usepackage{url}
\usepackage[usenames,dvipsnames,svgnames,table]{xcolor}
\usepackage[para,online,flushleft]{threeparttable}
\usepackage{cite}
\usepackage{balance}
\usepackage{times}

\usepackage{stfloats}
\usepackage{balance}

\usepackage{pifont}% http://ctan.org/pkg/pifont
\usepackage{tikz}
\newcommand*\circled[1]{\tikz[baseline=(char.base)]{
            \node[shape=circle,draw,inner sep=2pt] (char) {#1};}}
\begin{document}

\title{Multiple-Objective Packet Routing Optimization for Aeronautical {\it{ad-hoc}} Networks}

\author{Jiankang~Zhang,~\IEEEmembership{Senior Member,~IEEE},
        Dong~Liu,~\IEEEmembership{Member,~IEEE},
        Sheng~Chen,~\IEEEmembership{Fellow,~IEEE},
        Soon~Xin~Ng,~\IEEEmembership{Senior~Member,~IEEE},
        Robert~G.~Maunder,~\IEEEmembership{Senior~Member,~IEEE},
        Lajos~Hanzo,~\IEEEmembership{Life Fellow,~IEEE}
\thanks{J.~Zhang is with Department of Computing \& Informatics, Bournemouth University, BH12 5BB, U.K. (E-mail: jzhang3@bournemouth.ac.uk).} 
\thanks{D.~Liu is with the School of Cyber Science and Technology, Beihang
University, Beijing 100191, China (e-mail: dliu@buaa.edu.cn).}
\thanks{S.~Chen, S.~X.~Ng, R.~Maunder and L.~Hanzo are with School of Electronics and Computer Science, University of Southampton, SO17 1BJ, U.K. (E-mails: \{sqc,sxn,rm,lh\}@ecs.soton.ac.uk).} %
\thanks{The financial support of  the Engineering and Physical Sciences Research Council projects EP/W016605/1 and EP/P003990/1 (COALESCE) as well as of the European Research Council's Advanced Fellow Grant QuantCom (Grant No. 789028).} %
}

%\markboth{IEEE Transactions on Vehicular Technology,~Vol.~XX, No.~XX, XXX~2015}
{}
%{Shell \MakeLowercase{\textit{et al.}}: Bare Demo of IEEEtran.cls for Journals}

\maketitle

\begin{abstract}
Providing Internet service above the clouds is of ever-increasing interest and in this context  aeronautical {\it{ad-hoc}} networking (AANET) constitutes a promising solution. However, the optimization of packet routing in large ad hoc networks is quite challenging. In this paper, we develop a discrete $\epsilon$ multi-objective genetic algorithm ($\epsilon$-DMOGA) for jointly optimizing the end-to-end latency, the end-to-end spectral efficiency (SE), and the path expiration time (PET) that specifies how long the routing path can be relied on without re-optimizing the path. More specifically, a distance-based adaptive coding and modulation (ACM) scheme specifically designed for aeronautical communications is exploited for quantifying each link's achievable SE. Furthermore, the queueing delay at each node is also incorporated into the multiple-objective optimization metric. Our $\epsilon$-DMOGA assisted multiple-objective routing optimization is validated by real historical flight data collected over the Australian airspace on two selected representative dates.
\end{abstract}


\IEEEpeerreviewmaketitle

\section{Introduction}\label{S1}

Internet access has become almost ubiquitously supported by the global terrestrial mobile networks relying on the fourth-generation (4G) and fifth-generation (5G) wireless systems.  Hence, having Internet access has become virtually indispensable. The provision of Internet-above-the-clouds \cite{zhang2019aeronautical} is also of ever-increasing interest to both the civil aviation airlines and to the passengers.

In-flight WiFi relying on satellites and cellular systems has been available on the some flights of global airlines, such as British Airways, American Airlines, United Airlines, Emirates and Delta Airlines, just to name a few. However, aeronautical communications directly relying on satellites and/or cellular systems suffer from high cost, limited coverage, limited capacity, and/or high end-to-end latency. Furthermore, the cellular systems that can support ground-to-air (G2A) communications are limited to a line-of-sight range and require specially designed ground stations (GSs), which necessitate the roll-out of an extensive ground infrastructure to cover a wide area. Intuitively, it is quite a challenge to provide ubiquitous coverage for every flight at a low cost by directly relying on satellites and/or cellular systems. As an alternative architecture, aeronautical {\it{ad-hoc}} networks (AANETs) \cite{medina2011airborne,vey2014aeronautical,sakhaee2006aeronautical} are capable of extending the coverage, whilst reducing the communication cost by hopping messages from plane to plane. Each aircraft as a network node is capable of sending, receiving and relaying messages until the messages are delivered to or fetched from a GS, so as to enable Internet access.

Hence, routing, which finds an `\emph{optimal}' path consisting of a sequence of relay nodes, is one of the most important challenges to be solved in support of this Internet-above-the-clouds application. Routing protocols have been intensively investigated in mobile {\it{ad-hoc}} networks (MANETs) \cite{mauve2001survey} and vehicular {\it{{\it{ad-hoc}}}} networks (VANETs) \cite{li2007routing} as well as in the flying {\it{{\it{ad-hoc}}}} networks (FANETs) \cite{lakew2020routing}. However, as our analysis in \cite{zhang2019aeronautical} has revealed, AANETs have their unique features in terms of flying speed, altitude, propagation characteristics and network coverage as well as node mobility, which are different from those of MANETs, VANETs and FANETs. Therefore, the routing protocols specially developed for MANETs, VANETs and FANETs cannot be directly applied to AANETs, although their philosophies may be appropriately adopted. Hence, we will focus our attention on routing protocols specially designed for aeronautical networks.

Sakhaee and Jamalipour \cite{sakhaee2006global} showed that the probability of 
finding at least two but potentially up to dozens of aircraft capable of establishing an AANET above-the-cloud is close to $100\%$. It was inferred by investigating a snapshot of flight data over the United States (US). They also proposed a quality of service (QoS) based so-called multipath Doppler routing protocol by jointly considering both the QoS and the relative velocity of nodes in order to find stable routing paths. Luo {\it et al} \cite{luo2019AeroMRP} proposed a reliable multipath transport protocol for
AANETs by exploiting the path diversity provided by heterogeneous aeronautical networks. 
By exploiting the geographical information, Iordanakis {\it et al} \cite{iordanakis2006ad} proposed a routing protocol for aeronautical mobile ad hoc networks, which may be viewed as an evolved version of the classical {\it{ad-hoc}} on demand distance vector based routing (AODV) originally developed for MANETs. Furthermore, geographical information was intensively exploited in designing the routing protocols of \cite{jabbar2009aerorp,peters2011geographical,rohrer2011aerorp,medina2011geographic,gu2011delay,wang2013gr,mahmoud2013ads,swidan2015secure,zheng2016load,pang2017secure,luo2017multiple} for AANETs by considering that the locations of civil passenger  aircraft are always available with the aid of the on-board radar and the automatic dependent surveillance-broadcast (ADS-B) system \cite{ADS-B1,ADS-B2}. Explicitly, the authors of  \cite{jabbar2009aerorp,peters2011geographical,rohrer2011aerorp} developed a routing protocol termed as AeroRP, which is a highly adaptive location-aware routing algorithm exploiting the broadcast nature of the wireless medium along with the physical node location and trajectory knowledge for improving the data delivery in Mach-speed mobile scenarios. However, AeroRP ignores the delay imposed by relaying and it is prone to network congestion. Medina {\it et al.} \cite{medina2011geographic} proposed a geographic load sharing strategy for fully exploiting the total air-to-ground capacity available at any given instant in time. In their work, the network congestion was avoided by a congestion-aware handover strategy capable of efficient load balancing among Internet Gateways. Gu {\it et al.} \cite{gu2011delay} proposed a delay-aware routing scheme using a joint metric relying on both the relative velocity and the expected queueing delay of nodes for selecting the next node. Wang {\it et al.}  \cite{wang2020delay} also designed a delay-aware routing protocol for aeronautical networks, which explored the effect of dual connectivity on delay-aware resource control in heterogeneous aeronautical networks. Du {\it et al.} \cite{du2021dynamic} aimed for minimizing the
end-to-end transmission delay by jointly exploiting the direct transmissions,
relayed transmissions and opportunistic transmissions. By contrast, we have also minimized the end-to-end delay of AANETs by exploiting a weighted digraph and the shortest-path algorithm in \cite{cui2021minimum} and by invoking deep reinforcement learning in \cite{liu2021deep}, respectively.

Both Wang {\it et al.} \cite{wang2013gr} as well as Mahmoud and Larrieu \cite{mahmoud2013ads} exploited the geographical information provided by the ADS-B system. Specifically, Wang {\it et al.} \cite{wang2013gr} eliminated need for the traditional routing beaconing and improved the next hop selection, whilst Mahmoud and Larrieu \cite{mahmoud2013ads} concentrated on improving the information security of routing protocols. The security issues routing were further considered by Swidan {\it et al.} \cite{swidan2015secure} and Pang {\it et al.} \cite{pang2017secure}. Specifically, Swidan {\it et al.} \cite{swidan2015secure} proposed a secure geographical routing protocol by using the GS as a trusted third party for authentication and key transport. However, their solution required an additional transceiver at each aircraft having a communication range of 150\,km and wide downlink bandwidth for point-to-point communication with the GS. By contrast, Pang {\it et al.} \cite{pang2017secure} advocated an  identity-based public key cryptosystem, which relies on the authentication of neighbor nodes and establishes a shared secret during the neighbor discovery phase, followed by the encryption of the data during the data forwarding phase. Luo and Wang \cite{luo2017multiple} proposed a multiple QoS parameters-based routing protocol in order to improve the overall network performance for communication between aircraft and the ground. Explicitly, they jointly optimized the maximum path life-time period, maximum residual path load capacity, and minimum path latency from all the available paths between an aircraft node and the Internet gateways with the aid of carefully selected weighting factors for the path life-time period, residual path load capacity and path latency.

Since the overall network performance depends on multiple factors, it is unfair to optimize the overall network performance by relying on a single factor, such as the error probability, latency or capacity. As argued in \cite{zhang2019aeronautical}, in contrast to conventional single-objective optimization, multi-objective optimization is capable of finding all the global Pareto optimal solutions by potentially allowing the system to be reconfigured in any of it most desired optimal operating mode.  
In \cite{cui2021twin,cui2021multiobjective}, we developed a twin-component near-Pareto routing scheme by invoking the classic non-dominated Sorting Genetic Algorithm II (NSGA-II), which is capable of finding  a set of trade-off solutions in terms of the total delay and the throughput. Furthermore, in \cite{liu2021deeplearning}, we extended our single objective optimization efforts of \cite{liu2021deep} to multi-objective packet routing optimization in  AANETs in the north-Atlantic region. In the existing AANET literature, there is a paucity of contributions on applying multiple-objective routing optimization by jointly considering the end-to-end throughput, end-to-end latency and the corresponding path expiration time (PET).  However, the nodes in AANETs are airplanes, which typically fly at a speed of 880 to 926 km/h  \cite{zhang2019aeronautical}, hence a path may break quite soon. Hence, the PET also becomes a much more critical metric in AANETs compare to MANETs, VANETs and FANETs. Furthermore, most of the existing routing protocols were investigated mainly based on randomly generated flight data, which cannot reveal the network performance of real-world AANETs constituted by the real flights in the air. Against this background, we propose a multiple-objective routing optimization scheme for the AANET, and we evaluate its overall network performance using large-scale real historical flight data over the  Australian airspace. Explicitly, our main contributions are summarized as follows.
\begin{itemize}
\item [1)] We propose multi-objective routing optimization for jointly optimizing the end-to-end spectral efficiency (SE), the end-to-end latency and the PET. More specifically, the latency is addressed as one of the objectives by constraining the maximum number of affordable hops, whilst the congestion is addressed by imposing a certain queueing delay at each node. Furthermore, the distance-based adaptive coding and modulation (ACM) of \cite{zhang2017adaptive,zhang2018regularized}, which was specifically designed for aeronautical communications is adopted for quantifying the each link's SE so as to determine the final end-to-end SE. Naturally, the lowest link-throughput limits the entire path's throughput.
\item [2)] At the time of writing, there is no discrete $\epsilon$-MOGA version in the open literature and there is no application example of $\epsilon$-MOGA in the context of routing problems. Based on the philosophy of $\epsilon$ multi-objective genetic algorithm ($\epsilon$-MOGA) \cite{herrerowell}, which operates on a continuous parameter space for finding the Pareto-front of optimal solutions, we develop a discrete version of $\epsilon$-MOGA by considering the specific features of the routing paths consisting of discrete aircraft IDs, which we refer to as the discrete $\epsilon$ multi-objective genetic algorithm ($\epsilon$-DMOGA).  Explicitly, in order to accommodate the unique feature of routing problem in AANETs, we have adapted the existing $\epsilon$-MOGA to create a discrete $\epsilon$-MOGA by considering the discrete search space of routing problems in AANETs relying on  discrete aircraft IDs. This adaptation is not straightforward at all, because it involves new operations conceiving the encoding/decoding of chromosomes, as well as new crossover and mutation operations with respect to the specific nature of discrete variables that constitute a routing path emerging from a source aircraft node to a destination ground station. We use this $\epsilon$-DMOGA for efficiently solving the proposed multi-objective routing optimization problem.
\item [3)] The overall network performance of our multiple-objective routing optimization quantified in terms of the end-to-end latency, the end-to-end SE and the PET, are investigated based on large-scale real historical flight data recorded over the Australian airspace. More specifically, real historical flight data of two representative dates of the top-five airlines in Australia's domestic flights are exploited for our investigations.
\end{itemize}

The remainder of this paper is organized as follows. The network architecture is presented in Section~\ref{S2}, which includes the mobility model and the multiple-objective functions to be investigated. In Section~\ref{S3}, we develop a discrete version of $\epsilon$-MOGA, termed as $\epsilon$-DMOGA, by exploiting the discrete nature of the routing paths specified by the aircraft IDs, which provides an effective tool to solve our proposed multi-objective routing optimization. Our simulation results based on real historical flight data recorded over the Australian airspace are presented in Section~\ref{S4}, while our conclusions are offered in Section~\ref{S5}.

\begin{figure*}[tbp!]
\vspace{-2mm}
\begin{center}
  \includegraphics[width=0.75\textwidth]{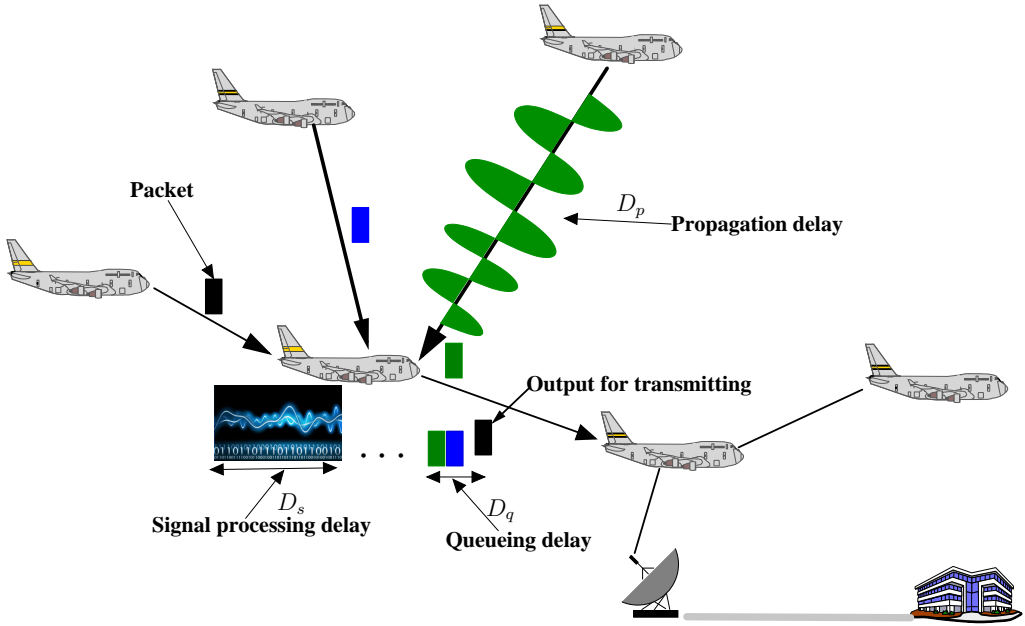} 
\end{center}
\vspace{-2mm}
\caption{Illustration for the sources of packet delay in routing}
\label{fig1A}
\vspace{-2mm}
\end{figure*}

\section{Network Architecture}\label{S2}

The avionic network considered takes into account the peculiarities of aeronautical communications and exploits them for optimizing our multiple objectives. Satellites are also included into the AANET considered, which are used as the last resort for an aircraft outside all the neighbouring aircrafts' communications range. In contrast to MANETs and VANETs, the nodes of an AANET are distributed in a 3D space, and they move at extremely high speed over long distances. The geographic information of each aircraft is available for routing optimization and network design, which can be obtained with the assistance of the global positioning system (GPS) and airborne radar carried by each aircraft \cite{batabyal2015mobility}. Moreover, ADS-B systems have been widely deployed on commercial passenger aircraft, which can also provide an information  vector including the aircraft ID, position, ground speed and heading directions \cite{luo2017multiple}.

\subsection{Mobility model}\label{S2-1}

In our network optimization, we consider all the aircraft during the 24 hours of both the busiest  and the quietest day in a year. Based on the historical flight observation on Flightradar24\footnote{Flightradar24 is a global flight tracking service that provides real-time information of aircraft around the world, which is accessible on \url{https://www.flightradar24.com}}, the busiest day was June 29th in 2018, whilst the quietest day was  December 25th in 2018. The movement of a node that represents an aircraft was recorded as real historical flight data, which typically includes the flight phases of holding, takeoff/landing, taxiing, and parking that always performs at an airport or near an airport, as well as the longest en-route phase. Contrast to the traditional nodes in MANETs and VANETs as well as even the nodes in FANETs, the nodes in AANETs move at a relatively high speed during the en-route phase, typically at velocities of 800 to 1000\,km/h. 

As our evaluation is based on real historical flight data over the Australian airspace, the nodes tend to be sparser compared to Europe and North-America. The nodes typically fly between coastal cities, such as Sydney, Melbourne, Perth and Gold Coast. The eastern coastal airspace is much busier than the northern, western and southern coastal airspace, since most people reside in eastern coastal cities, and most international flights depart from/arrive at eastern coastal cities. The central Australian airspace quite sparse, since only few people live in the central area of Australia.

\subsection{Objective functions}\label{S2-2}

In the Australian aeronautical network, there are $N_{G}$ GSs at five Australian airports. An aircraft can access the Internet by relying on an optimal routing path to a GS. Hence the flights do not have to rely on costly satellites as a relay node to access the Internet. 

We will consider the achievable end-to-end latency and end-to-end throughput as well as the stability of the routing path. Explicitly, the end-to-end latency is the sum of the signal propagation delays, signal processing delays and queuing delays. The end-to-end throughput is determined by the specific link having the minimum throughput. Again, the stability of the routing path is quantified in terms of the PET, which is in turn determined by the link having the minimum expiration time given a specific ACM mode.

\subsubsection{The end-to-end latency}

Let us now quantify the propagation, signal processing and queuing delays.  Although there may be a ground-reflection component in the received signal of aeronautical communications, it is dominated by the Line of Sight (LoS) path for air-to-air (A2A) communications in AANETs \cite{zhang2017adaptive}. Hence, the ground-reflected component may be neglected in A2A transmission and the propagation delay can be modelled by that of the LoS path limited by the speed of light. Explicitly,  let the distance between node $r_{n}$ and node $r_{n + 1}$ be denoted by  $d_{r_{n},r_{n + 1}}$. The propagation delay $D_{p}(r_{n} \rightarrow r_{n + 1})$, as illustrated in Fig.~\ref{fig1A}, between node $r_{n}$ and node $r_{n + 1}$ is given  by
\begin{align}\label{eq1}
D_{p}(r_{n} \rightarrow r_{n + 1}) = \frac{d_{r_{n},r_{n + 1}}}{c} ,
\end{align}
where $c$ is the speed of light, $r_{n}\! \rightarrow\! r_{n + 1}$ is the link spanning from node $r_{n}$ to node $r_{n + 1}$ in the routing path $\bm{r}\! =\! \{r_{n}\! \rightarrow\! r_{n + 1}\! \rightarrow\! \cdots\! \rightarrow\! r_{N + n - 1}\}$, which consists of $N\! -\! 1$ hops from the source node $r_n$ to the destination node $r_{n + N - 1}$, with $N$ being the number of nodes in the routing path $\bm{r}$.

As shown in Fig.~\ref{fig1A}, the signal processing delay is the time that a relay node takes to process a packet before it can forward it to its output queue, which includes the decoding and forwarding operations, destination lookup, packet-header updates, etc. Intuitively, when node $r_n$ is the source node, obviously there is no  decoding and forwarding operations, destination lookup, packet-header updates, hence $D_s(r_n \to r_{n+1}) = 0$ in this case. However, when node $r_n$ is not a source node, but a relay node, it has to carry out the  decoding and forwarding operations, destination lookup, packet-header updates. Then a constant signal processing delay will be imposed. Depending on the digital signal processing capability of the hardware and the detailed operations needed, the signal processing latency ranges from 0.5 to ten milliseconds with some complex designs having as much as 30 milliseconds \cite{McNell2021networked}. In our investigations, we set it to 5 ms as a compromise consideration. Without loss of generality, we formulate signal processing delay at node $r_{n}$ as 
\begin{align}\label{eq2}
D_{s}(r_{n} \rightarrow r_{n + 1})  =  \left\{
\begin{array}{ll}
0\,\text{ms}, & \text{if node $r_{n}$ is the source node} , \\
5\,\text{ms}, & \text{if node $r_{n}$ is not a source node} .
\end{array}
\right.
\end{align}

The queueing delay is the time that a packet waits in a relay node after arriving at the node's queue until it can be processed plus the waiting time before the processed packet can be transmitted to the output link, as illustrated in Fig.~\ref{fig1A}. The input  queueing delay is proportional to the number of packets that have already been waiting in the queue at a given time, while the output transmission delay is upper bounded by the time needed for transmitting a packet. Extensive research has been dedicated to queue theory, queue scheduling and/or minimizing the queuing delay at a node in networks \cite{das2015spatial,baz2014analysis,al2019queuing}. We can reasonably model the input queueing delay at node $r_n$ as follows
\begin{align}\label{eq3}
D_{q_{1}} = O_{r_{n}}D_{q_{0}} ,
\end{align}
where the indicator $O_{r_n}\! \in\! \{0,1,\cdots,N_{B}\}$, which indicates how many packets are at the front of the queue, and $N_{B}$ is the maximum number of packets that the node can have in its queue, while $D_{q_{0}}$ is a fixed processing delay related to transmitting a whole packet through the node's output link. The total buffering delay, denoted as $D_{q_2}$, is clearly upper bounded by $D_{q_0}$, i.e., we have $D_{q_2}\! \le\! D_{q_0}$. The total queueing delay of forwarding a packet from $r_n$ to $r_{n+1}$, which is the sum of the input queueing delay $D_{q_1}$ and the output buffering delay $D_{q_2}$, can be expressed as
\begin{align}\label{eq4}
D_{q}(r_{n} \rightarrow r_{n + 1}) = D_{q_{1}} + D_{q_{2}} \approx \left(O_{r_n} + 1\right)D_{q_{0}} .
\end{align}
Note that a packet can only be routed through a node if the node's queue is not full, i.e. the number of packets waiting in the node's queue is less than $N_B$. This imposes a constraint on the routing decisions. Again, the delay imposed on a packet during its passage from node $r_{n}$ to node $r_{n + 1}$ is the sum of the signal propagation delay, signal processing delay and queuing delay, which is given by
\begin{align}\label{eq5}
D(r_{n} \rightarrow r_{n + 1}) =& D_{p}(r_{n} \rightarrow r_{n + 1})+ D_{s}(r_{n} \rightarrow r_{n + 1}) \nonumber \\
& + D_{q}(r_{n} \rightarrow r_{n + 1}) .
\end{align}
Therefore, the end-to-end latency of the routing path $\bm{r}$ is given by
\begin{align}\label{eq6}
D(\bm{r}) = \sum\limits_{n = 1}^{N - 1}D(r_{n} \rightarrow r_{n + 1}) .
\end{align}

\subsubsection{The end-to-end throughput}

The end-to-end throughput is determined by the link in the routing path $\bm{r}$, which has the minimum link throughput. Let $C(r_{n} \rightarrow r_{n + 1})$ denote the link throughput between node $r_{n}$ and node $r_{n + 1}$. Then the end-to-end throughput is given by
\begin{align}\label{eq7}
C(\bm{r}) = \min\limits_{1\le n\le N-1} C(r_{n} \rightarrow r_{n + 1}).
\end{align}
The achievable link throughput $C(r_{n} \rightarrow r_{n + 1})$ is affected by the  channel conditions and other factors, such as the co-channel inference. The link throughput is  a function of instantaneous ignal-to-interference-plus-noise
ratio (SINR), where the instantaneous SINR may be estimated using pilot signals in traditional terrestrial mobile communications. However, the problem  in aeronautical communication applications is that the high speed of aircraft may result in uncorrelated small-scale fading and consequently in unreliable estimates of the instantaneous SNR  or SINR, further aggravated by frequently switching among the ACM modes. Using erroneous instantaneous  SNR or SINR estimates for frequently switching ACM modes may cause frequent unsuccessful transmissions, because the SINR estimates quickly become obsolete. Moreover, the instantaneous SINR does fluctuate around its average, but it has a limited range in the typical LoS scenarios. Hence, 
in the routing problem in AANETs, the best available distance-based ACM \cite{zhang2017adaptive,zhang2018regularized} is invoked for quantifying the link throughput of the air-to-air aeronautical communications. Hence, the achievable link throughput $C(r_{n} \rightarrow r_{n + 1})$ is a function of the communication distance between node $r_{n}$ and node $r_{n + 1}$, which is also affected by the co-channel interference imposed by the neighbour aircraft. Furthermore, given a set of $K$ ACM modes, there are $K\! +\! 1$ distance thresholds $\{d_{k}\}_{k=0}^{K}$. The data-transmitting aircraft selects an ACM mode to transmit/relay the data according to
\begin{align}\label{eq8}
\text{if}\quad d_{k} \le d < d_{k-1} \quad \text{choose $k$-th ACM mode} ,
\end{align}
where $d_0\! =\! D_{\max}\! =\! D_{\text{A2A}}\! >\! d_1\! >\! \cdots\! >\! d_{K-1}\! >\! d_K\! =\! D_{\min}$. Clearly, if the distance $d$ is outside the range of $[D_{\text{min}},~D_{\text{max}}]$, there will be no adequate communication link. More specifically, $D_{\text{min}}$ is the minimum flight-safety separation that must be obeyed according to the aviation safety regulation, whilst $D_{\text{max}}$ is the maximum communication range of two aircraft, which is given by the radio distance to horizon of A2A communication \cite{zhang2019aeronautical}.

\subsubsection{The path expiration time}

The PET is determined by the most vulnerable link of the routing path $\bm{r}$ that has the shortest link expiration time (LET). Let $T_{\text{LET}}(r_{n} \rightarrow r_{n + 1})$ be the LET of the link between node $r_{n}$ and node $r_{n + 1}$ for offering ACM mode-$i$. Then the PET of $\bm{r}$ is given by
\begin{align}\label{eq9}
T_{\text{PET}}(\bm{r}) = \min\limits_{1\le n\le N-1} T_{\text{LET}}(r_{n} \rightarrow r_{n + 1}) .
\end{align}
Since the ACM is adopted, we have to modify the formulation of calculating the LET given in \cite{sakhaee2006global,luo2017multiple}. Specifically, given the speeds $v_{r_{n}}$ and $v_{r_{n + 1}}$, the heading directions $\theta_{r_{n}}$ and $\theta_{r_{n + 1}}$ as well as the coordinates $(x_{r_{n}}, y_{r_{n}})$ and $(x_{r_{n + 1}}, y_{r_{n + 1}})$ for node $r_{n}$ and node $r_{n + 1}$, respectively, assume that the distance between $r_{n}$ and $r_{n + 1}$ satisfies $d_k \le d_{r_{n},r_{n + 1}} < d_{k-1}$, where $d_{k-1}$ is the distance threshold or maximum distance that may be bridged over by the ACM mode~$k$, as defined in Eq.\,(\ref{eq8}). Then the LET of $T_{\text{LET}}(r_{n} \rightarrow r_{n + 1})$ is calculated according to:
\begin{align}\label{eq10}
T_{\text{LET}}(r_{n} \!\!\rightarrow \!\!r_{n + 1})\!\! = \!\!\frac{(ab \!\!+\!\! ef)\!\! +\!\! \sqrt{(a^2 \!\!+\!\! e^2)d_{k-1}^2 \!\!-\!\! (af\!\! -\!\!be)^{2}}}{a^{2} \!\!+ \!\!e^{2}} ,
\end{align}
where $a, b, e$ and $f$ are given by \cite{sakhaee2006global,luo2017multiple}
\begin{align} % eqs.11-14
a &= v_{r_{n}}\cos \theta_{r_{n}} - v_{r_{n + 1}}\cos \theta_{r_{n + 1}} , \\
b &= x_{r_{n}} - x_{r_{n + 1}} ,\\
e &= v_{r_{n}}\sin \theta_{r_{n}} - v_{r_{n + 1}}\sin \theta_{r_{n + 1}} ,\\
f &= y_{r_{n}} - y_{r_{n + 1}}  .
\end{align}
Intuitively, when aircraft $r_{n}$ and aircraft $r_{n + 1}$ have the same speed and heading direction, the LET between them is theoretically infinity. However, the associated LET is always upper-bounded by their flight time in practice. When aircraft $r_{n}$ and aircraft $r_{n + 1}$ have the exact opposite heading direction, they will have the minimum LET.

\subsection{Multi-objective routing optimization}\label{S2-3}

The specific multi-objective optimization is advocated here aims for maximizing the end-to-end achievable throughput, for minimizing the end-to-end latency and for maximizing the PET, which is formulated as
\begin{align}\label{eq15}
\left\{
\begin{array}{l}
\mathcal{J}_{1}(\bm{r}) = \max \,C(\bm{r}) , \\
\mathcal{J}_{2}(\bm{r}) = \min \,D(\bm{r}) , \\
\mathcal{J}_{3}(\bm{r}) = \max \,T_{\text{PET}}(\bm{r}) ,
\end{array}
\right.
\end{align}
\begin{align}\label{eq16}
\text{s.t.} 
\left\{
\begin{array}{l}
 D(\bm{r}) \le 250\,\text{ms} , \\
 N - 1 \le 5 .
\end{array}
\right.
\end{align}
The round-trip latency of  geostationary satellite links is about 250 ms \cite{medina2011geographic},  which is imposed by the propagation delay up and down from the satellite. Hence intuitively, the end-to-end latency should be less than 250 ms, which results in the first constraint $D(\bm{r}) \le 250\,\text{ms}$. Naturally, when low earth-obit (LEO) satellites are considered at say 600 Km altitude, their round-trip delay is as low as 4 ms. The second constraint of $N - 1 \le 5$ is based on a practical consideration of the Australian scenario. Explicitly, the aircraft tend to fly over the land of Australia, where the GSs are at airports on the ground. Hence, an aircraft is typically capable of accessing the Internet with a small number of hops. Furthermore, the second constraint of $N - 1 \le 5$ is also used for limiting the number of nodes involved, which is helpful for controlling the AANET size. Since a routing path having more hops is more vulnerable to cyber attacks, limiting the number of nodes in a routing path also helps to secure information transmission. Clearly, for geographic areas, such as the AANET over the Atlantic Ocean, we will have to set a higher value for the maximum number of hops.

\section{Discrete $\epsilon$-MOGA based Pareto-optimization of AANET routing problem}\label{S3}

No closed-form solution can be derived for the multi-objective optimization problem (\ref{eq15}) under the constraint (\ref{eq16}). There are diverse methods of solving multi-objective optimization problems, such as the Lexicographic method \cite{Isermann1982linear}, weighted sum method \cite{Zadeh1963Optimality}, elitist non-dominated sorting genetic algorithm (NSGA-II) \cite{Zitzler1999Multiobjective}, Strength Pareto Evolutionary Algorithm (SPEA) \cite{Deb2002afast}, SPEA-II \cite{Zitzler2002SPEA2}, Pareto Enveloped based Selection Algorithm (PESA) \cite{Corne2000the} and PESA-II \cite{Corne2001PESA-II} as well as numerous other variations with their pros and cons. For example, the Lexicographic method is sensitive to the iteration order of objectives, while the weighted method is sensitive to the weightings and both of them suffer from high computational burden. The NSGA-II does not perform very well for more objectives, while the SPEA and SPEA2 are also exhibit high computational complexity, and the PESA as well as  PESA2 are sensitive to the size of hyperbox. By contrast, as a member of the elitist multi-objective evolutionary algorithm family based on the concept of $\epsilon$-dominance, $\epsilon$-MOGA has the compelling characteristics of efficient parallel computing along with the efficient control of the elitist archive, where problem solutions are stored. Hence, $\epsilon$-MOGA outperforms the above-mentioned multi-objective optimization algorithms in terms of its  convergence, diversity of solutions and computational efficiency.  Hence, we apply the $\epsilon$-DMOGA for determining the optimal Pareto-front at a moderate computational burden. Our $\epsilon$-DMOGA is developed from the $\epsilon$-MOGA \cite{herrerowell,Martinez2009Applied}, which is an elitist multi-objective evolutionary algorithm based on the concept of $\epsilon$-dominance \cite{reynoso2014controller}, by taking into consideration the discrete nature of the routing path constituted by discrete aircraft IDs.

\subsection{$\epsilon$-DMOGA}
In the Pareto optimal set for the multi-objective optimization problem (\ref{eq15}), no single  solution should dominate others. Explicitly, a solution routing path $\bm{r}_{1}$ by definition dominates another routing path $\bm{r}_{2}$ in the routing path space, if and only if all the objectives of $\bm{r}_{1}$ are no worse than the objectives of $\bm{r}_{2}$ and at least one objective of $\bm{r}_{1}$ is better than that of $\bm{r}_{2}$, which is formulated as
\begin{align}\label{eq17}
\forall i \!\!=  \!\!1,2,3, \mathcal{J}_{i}(\bm{r}_{1})  \!\!\preceq  \!\!\mathcal{J}_{i}(\bm{r}_{2}) \,\, \text{and}\,\, \exists k  \!\!=  \!\!1,2,3, \mathcal{J}_k(\bm{r}_{1})  \!\!\prec \!\! \mathcal{J}_k(\bm{r}_{2}) .
\end{align}
The operator $\preceq$ represents that the lefthand objective is no worse than the righthand one. For example, $\mathcal{J}_{1}(\bm{r}_{1}) \preceq \mathcal{J}_{1}(\bm{r}_{2})$ is equal to $C(\bm{r}_{1}) \ge C(\bm{r}_{2})$, whilst $\mathcal{J}_{2}(\bm{r}_{1}) \preceq \mathcal{J}_{2}(\bm{r}_{2})$ is equal to $D(\bm{r}_{1}) \le D(\bm{r}_{2})$. Similarly, $\prec$ represents that the lefthand objective is better than the righthand one. Then, the Pareto-front solution set $\mathbf{R}$ can be formulated as
\begin{align}\label{eq18}
\mathbf{R} = \left\{\bm{r} \in \mathbf{R}|\nexists \widetilde{\bm{r}} \in \mathbf{R}\,: \,\widetilde{\bm{r}} \preceq \bm{r}\right\} ,
\end{align}
where $\nexists$ represents `does not exist'. Hence, for $\bm{r}\in \mathbf{R}$, $\nexists \widetilde{\bm{r}} \in \mathbf{R}\,: \,\widetilde{\bm{r}} \preceq \bm{r}$ means that there is no single $\widetilde{\bm{r}}$, which dominates $\bm{r}$ if it does not belong to $\mathbf{R}$.

Again, without loss of generality, we discuss the AANET over the Australian airspace. For other AANETs, similar discussions can be applied subject to minor modifications. Intuitively, the end-to-end throughput is a more dominant criterion for the Internet-above-the-clouds than the end-to-end latency and the PET. Thus, our $\epsilon$-DMOGA based multi-objective routing optimization may start from a direct connection to any of the $N_G$ GSs by finding the Pareto-front of optimal solutions with respect to throughput, latency and PET. Since $N_G$ is small, we can easily find a `Pareto-optimal' GS that dominates other GSs by enumerating the multiple objectives to each GS. This is effectively the single-hop solution. Then, the $\epsilon$-DMOGA based multi-objective routing optimization can proceed to find all the Pareto-front solutions within an affordable number of hops, say $N-1=2, 3, 4, 5$ with respect to the multi-component objective function (\ref{eq15}).  

The $\epsilon$-DMOGA based multi-objective routing optimization is characterized by its initialization, individual mutation, crossover and selection operations used throughout exploring the search space in a generation-based progression, until the termination criterion is met. We now detail this $\epsilon$-DMOGA.

\begin{itemize}
\item [1)] \textbf{Initialization}. At the first generation of $g=1$, where $g$ denotes the generation index, the $\epsilon$-DMOGA commences its search by randomly generating an initial population of $P_s$ $N$-element routing path vectors, denoted as $\bm{P}^{(g)}$. Explicitly, permutation encoding is invoked for generating a chromosome representing a routing path, which consists of a string of aircraft IDs from source aircraft to the destination ground station. For each individual of $\bm{P}^{(g)}$, the first element is the source node $r_1$, and its second element is randomly selected from the node space 
\begin{align}\label{eq19}
\mathbb{A}\backslash\left\{r_1\right\} = \left\{r \in \mathbb{A}, \sim \left(r \in \{r_1\}\right)\right\} ,
\end{align}
where $\mathbb{A}$ is the node space consisting of the aircraft in air, and $\mathbb{A}\backslash\{r_1\}$ represents the node space with $r_1$ removed. In general, the $n$-th element of an individual, where $2\le n\le N-1$, is randomly selected from the node space 
\begin{align}\label{eq20}
&\mathbb{A}\backslash\left\{r_{1},r_{2},\cdots,r_{n - 1}\right\} \nonumber\\
&\qquad= \left\{r \in \mathbb{A}, \sim \left(r \in \{r_{1},r_{2},\cdots,r_{n - 1}\right\}\right)\} .
\end{align}
The last node in a routing path, i.e. the $N$-th node, is a GS, randomly selected from the GS node space $\mathbb{B}$, which consists of the $N_G$ GSs at airports. The archive $\bm{A}^{(g)}$ that contains the elite population is initialized as the null set at the first generation $g=1$.

The $\epsilon$-DMOGA scheme solves the multi-objective routing optimization by evolving the main population $\bm{P}^{(g)}$ of $P_s$ $N$-element routing path vectors from one generation to the next.

\begin{figure}[tbp!]
\vspace{-2mm}
\begin{center}
  \includegraphics[width=0.75\columnwidth]{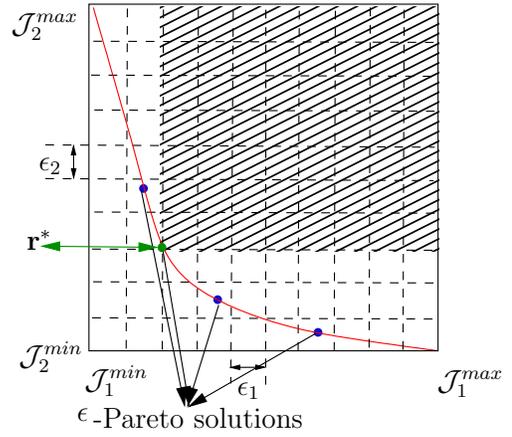} 
\end{center}
\vspace{-2mm}
\caption{The illustration of $\epsilon$-dominance and the $\epsilon$-Pareto-front solution.}
\label{fig1}
\vspace{-2mm}
\end{figure}

\item [2)] \textbf{Archive}. By calculating and comparing the multi-objective functions for the individuals of  $\bm{P}^{(g)}$, the $\epsilon$-Pareto-front solution set $\widetilde{\mathbf{R}}$ is selected. Explicitly, the individuals in the $\epsilon$-Pareto-front solution set $\widetilde{\mathbf{R}}$ $\epsilon$-dominate all the other individuals that are not selected into $\widetilde{\mathbf{R}}$. The concept of $\epsilon$-dominance is illustrated in Fig.~\ref{fig1}\footnote{In Fig.~\ref{fig1}, we only plot two dimensions in order to have a clear illustration. For our three objective dimensions as formulated in (\ref{eq15}), the light blue areas will be cubic.}, where the shade areas are $\epsilon$-dominated by $\bm{r}^{*}$ and the other blue points on the $\epsilon$-Pareto-front are $\epsilon$-Pareto-front solutions. Furthermore, $\epsilon_i$, $i = 1,2,3$, is the width of a box, which is defined as
\begin{align}\label{eq21}
\epsilon_i = \frac{\mathcal{J}_{i}^{max} - \mathcal{J}_{i}^{min}}{N_{\text{box},i}} ,
\end{align}
where $N_{\text{box},i}$ is the number of partitions in the dimension of the $i$-th objective, which preserves the diversity of $\epsilon$-dominance solution in the $i$-th objective dimension. The Pareto-front limits $\mathcal{J}_{i}^{min}$ and $\mathcal{J}_{i}^{max}$ for $i = 1,2,3$ are calculated as follows
\begin{align}\label{eq22}
\mathcal{J}_{i}^{max} = & \max\limits_{\bm{r} \in \widetilde{\mathbf{R}}} \mathcal{J}_{i}\left(\bm{r}\right), i = 1,2,3 ,\\
\label{eq23}
\mathcal{J}_{i}^{min} = & \min\limits_{\bm{r} \in \widetilde{\mathbf{R}}} \mathcal{J}_{i}\left(\bm{r}\right), i = 1,2,3. 
\end{align}
The individuals in $\widetilde{\mathbf{R}}$ that are not $\epsilon$-dominated by the individuals in the elite population archive $\bm{A}^{(g)}$ are stored into $\bm{A}^{(g)}$. Hence the size of the archive $\bm{A}^{(g)}$ may vary at different generations. Furthermore, $\widetilde{\mathbf{R}}$ is an intermediate Pareto-front set at the current generation that converges towards the Pareto optimal set $\mathbf{R}$ as the population evolves over generations. 

\item [3)] \textbf{Variant}. A new variant is generated by the amalgamation of the `\emph{crossover}' and `\emph{mutation}' operations, which are typically two separate operations in single-objective GA optimization. However, we  use the terminology `variant' for the operations of crossover and mutation in our $\epsilon$-DMOGA, since both operations are controlled by the probability of crossover/mutation $p_{c/m}$. Explicitly, two individuals, $\bm{r}^{(P)}$ and $\bm{r}^{(A)}$, are randomly selected from the main population $\bm{P}^{(g)}$ and the elite population $\bm{A}^{(g)}$, respectively. Then, a randomly generated value $\alpha\in [0, ~ 1]$ decides which operation should be applied to $\bm{r}^{(g,P)}$ and $\bm{r}^{(g,A)}$.

\begin{itemize}
\item [\circled{1}] {\bf{Crossover}}. If $\alpha > p_{c/m}$, $\bm{r}^{(g,P)} = \left\{r_{1}^{(g,P)},r_{2}^{(g,P)},\cdots,r_{N}^{(g,P)}\right\}$ and $\bm{r}^{(g,A)} = \Big\{r_{1}^{(g,A)},r_{2}^{(g,A)},$ $ \cdots,r_{N}^{(g,A)}\Big\}$ will cross over part of their elements. There exist numerous crossover mechanisms, and we opt for employing
the single-point crossover due to its simplicity. Explicitly, a point on both $\bm{r}^{(g,P)}$ and $\bm{r}^{(g,A)}$ is picked randomly, which is designated as the crossover point. Then the elements to the right of the crossover point are swapped between $\bm{r}^{(P)}$ and $\bm{r}^{(A)}$, which results in two new offspring, each carrying some genetic information from both parents. Given the crossover point $n\ge 2$, the two new offspring can be expressed as
\begin{align}\label{eq24}
\left\{ \begin{array}{lll}
\!\!\!\!\widehat{\bm{r}}_1^{(g,G)} \!\!\!\!\!\!&=&\!\!\!\!\!\! \left\{\!\!r_{1}^{(g,P)},r_{2}^{(g,P)},\!\!\cdots,\!\!r_{n}^{(g,P)},r_{n + 1}^{(g,A)}\!\!\cdots,\!\!r_{N}^{(g,A)}\!\!\right\} ,\\
\!\!\!\!\widehat{\bm{r}}_2^{(g,G)} \!\!\!\!\!\!&=&\!\!\!\! \!\!\left\{\!\!r_{1}^{(g,A)},r_{2}^{(g,A)},\!\!\cdots,\!\!r_{n}^{(g,A)},r_{n + 1}^{(g,P)}\!\!\cdots,\!\!r_{N}^{(g,P)}\!\!\right\} ,
\end{array} \right.
\end{align}
where the superscript $G$ indicates that both $\widehat{\bm{r}}_1^{(g,G)}$ and $\widehat{\bm{r}}_2^{(g,G)}$ are stored into an auxiliary population $\bm{G}^{(g)}$. Note that the same aircraft ID should be avoided both in $\widehat{\bm{r}}_1^{(g,G)}$ and $\widehat{\bm{r}}_2^{(g,G)}$ by applying sophisticated operations, such as for example, checking and mutating to a new non-same aircraft ID if a $r_{n + j}^{(g,A)}$ is same as a $r_{j}^{(g,P)}$.

\begin{figure}[tbp!]
\vspace{-4mm}
\begin{center}
  \includegraphics[width=0.75\columnwidth]{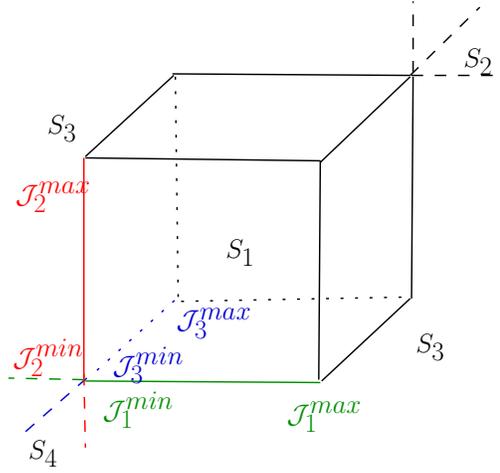} 
\end{center}
\vspace{-2mm}
\caption{The objective space can be divided into the four areas, namely, $S_1$: $\mathcal{J}_i^{min}\le \mathcal{J}_i\le\mathcal{J}_i^{max}$, $\forall i = 1,2,3$; $S_2$: $\mathcal{J}_i > \mathcal{J}_i^{max}$, $\forall i = 1,2,3$; $S_4$: $\mathcal{J}_i< \mathcal{J}_i^{min}$, $\forall i = 1,2,3$; and $S_3$: the rest of the objective space.}
\label{fig2}
\vspace{-2mm}
\end{figure}

\begin{figure*}[bp!]\setcounter{equation}{24}
\hrulefill
\vspace*{-1mm}
\begin{align}\label{eq25}
\left\{\begin{array}{lll}
\widehat{\bm{r}}_1^{(g,G)} &=& \left\{r_{1}^{(g,P)},\cdots,\widehat{r}_{l_1}^{(g,P)},\cdots,\widehat{r}_{l_2}^{(g,P)},\cdots,\widehat{r}_{l_{N_m}}^{(g,P)},\cdots,r_{N}^{(g,P)}\right\} ,\\
\widehat{\bm{r}}_2^{(g,G)} &=& \left\{r_{1}^{(g,A)},\cdots,\widehat{r}_{l_1}^{(g,A)},\cdots,\widehat{r}_{l_2}^{(g,A)},\cdots,\widehat{r}_{l_{N_m}}^{(g,A)},\cdots,r_{N}^{(g,A)}\right\} ,
\end{array}\right.
\end{align}
\vspace*{-1mm}
\end{figure*}

\item [\circled{2}] {\bf{Mutation}}. If $\alpha \le p_{c/m}$, $\bm{r}^{(g,P)} = \left\{r_{1}^{(g,P)},r_{2}^{(g,P)},\cdots,r_{N}^{(g,P)}\right\}$ and $\bm{r}^{(g,A)} = \Big\{r_{1}^{(g,A)},r_{2}^{(g,A)},$ $ \cdots,r_{N}^{(g,A)}\Big\}$ will mutate some of their elements. Intuitively, the mutation may occur in a single element or in multiple elements. We opt for the latter. Explicitly, an integer $N_m$ is randomly generated in the range of $[1, ~ N - 1]$. Then an $N_m$-length vector $\bm{l} = \left\{l_1,l_2,\cdots,l_{N_m}\right\}$ is generated and each of its elements $l_i$, $i = 1,2,\cdots, N_m$, is selected from the integer set of $\left\{2,3,\cdots,N\right\}$ without repetition. Specifically, $N_m$ determines the number of mutated elements with $l_i$, $1\le i\le N_m$, specifying the positions of these elements. The pair of new offspring generated by mutation are expressed as Eq.~(\ref{eq25}), 
%\begin{align}\label{eq25}
%\left\{\begin{array}{lll}
%\widehat{\bm{r}}_1^{(g,G)} &=& \left\{r_{1}^{(g,P)},\cdots,\widehat{r}_{l_1}^{(g,P)},\cdots,\widehat{r}_{l_2}^{(g,P)},\cdots,\widehat{r}_{l_{N_m}}^{(g,P)},\cdots,r_{N}^{(g,P)}\right\} ,\\
%\widehat{\bm{r}}_2^{(g,G)} &=& \left\{r_{1}^{(g,A)},\cdots,\widehat{r}_{l_1}^{(g,A)},\cdots,\widehat{r}_{l_2}^{(g,A)},\cdots,\widehat{r}_{l_{N_m}}^{(g,A)},\cdots,r_{N}^{(g,A)}\right\} ,
%\end{array}\right.
%\end{align}
where $\widehat{r}_{l_i}^{(g,P)}$ and $\widehat{r}_{l_i}^{(g,A)}$, $i = 1,2,\cdots,N_m$, are the new genes generated by mutation from parent $\bm{r}^{(g,P)}$ and $\bm{r}^{(g,A)}$, respectively. These mutated elements are randomly drawn from the aircraft node set. However, the mutated elements must not be duplicated with other elements within the same individual to avoid loops in the routing path. Therefore, if a mutated element is the same as another element in the same individual, it must be mutated again until it becomes different. Similarly, both $\widehat{\bm{r}}_1^{(g,G)}$ and $\widehat{\bm{r}}_2^{(g,G)}$ are stored into the auxiliary population $\bm{G}^{(g)}$.
\end{itemize}

The crossover or mutation operations are operated $N_O / 2$ times, which results in total of $N_O$ new offspring in the auxiliary population $\bm{G}^{(g)}$.

\item [4)] \textbf{Selection}. The selection operation of multiple-objective optimization is much more complex than that of single-objective optimization. Explicitly, the $\epsilon$-DMOGA calculates the multiple objective functions of the individuals in the auxiliary population $\bm{G}^{(g)}$ and decides,  which specific individual will be selected into the elite population $\bm{A}^{(g)}$ on the basis of its location in the objective space, as illustrated in Fig.~\ref{fig2}. More specifically, there are four scenarios depending on the particular location of the individual in the objective space.

\begin{itemize}
\item [\circled{1}] {\emph{Located in $S_1$}}. If an individual $\widehat{\bm{r}}_i^{(g,G)}$, $i\in\{1,2,\cdots, N_O\}$ is located in the objective function space area $S_1$ and it is not $\epsilon$-dominated by any individual of $\bm{A}^{(g)}$, it will be stored into the elite population $\bm{A}^{(g)}$, and the individuals in $\bm{A}^{(g)}$ that are $\epsilon$-dominated by $\widehat{\bm{r}}_i^{(g,G)}$ will be removed from the elite population.

\item [\circled{2}] {\emph{Located in $S_2$}}. If an individual $\widehat{\bm{r}}_{i}^{(g,G)}$, $i\in\{1,2,\cdots, N_O\}$ is located in the objective function space area $S_2$, it will not be stored into the elite population $\bm{A}^{(g)}$, since it is $\epsilon$-dominated by all the individuals in $\bm{A}^{(g)}$.

\item [\circled{3}] {\emph{Located in $S_3$}}. If an individual $\widehat{\bm{r}}_i^{(g,G)}$, $i\in\{1,2,\cdots, N_O\}$ is located in the objective function space area $S_3$, the $\epsilon$-DMOGA calculates and compares the objective functions of the individuals in $\widetilde{\bm{P}}^{(g)} = \bm{A}^{(g)} \cup \widehat{\bm{r}}_i^{(g,G)}$. Then, the $\epsilon$-Pareto-front set $\widetilde{\mathbf{R}}^{(\widetilde{\mathbf{P}})}$ is selected and the elite population $\bm{A}^{(g)}$ is updated as $\widetilde{\mathbf{R}}^{(\widetilde{\mathbf{P}})}$. Additionally, both the Pareto-front limits $\mathcal{J}_i^{min}$ and $\mathcal{J}_i^{max}$ as well as the box width $\epsilon_i$  are updated for all the three dimensions $i = 1,2,3$ according to (\ref{eq23}), (\ref{eq22}) and (\ref{eq21}).

\item [\circled{4}] {\emph{Located in $S_4$}}. If an individual $\widehat{\bm{r}}_{i}^{(g,G)}$, $i\in\{1,2,\cdots, N_O\}$ is located in the objective function space area $S_4$, all the individuals in the elite population $\bm{A}^{(g)}$ are deleted, since all of them are $\epsilon$-dominated by $\widehat{\bm{r}}_1^{(g,G)}$, and $\widehat{\bm{r}}_1^{(g,G)}$ is stored into $\bm{A}^{(g)}$. The limit of each objective function $\mathcal{J}_i^{min}$, $i = 1,2,3$, is updated as $\mathcal{J}_i\left(\widehat{\bm{r}}_1^{(g,G)}\right)$, $i = 1,2,3$.
\end{itemize}

\item [5)] \textbf{Update}. Update the main population $\bm{P}^{(g)}$ by comparing its individuals and the individuals selected from the auxiliary population $\bm{G}^{(g)}$. Explicitly, an individual $\widehat{\bm{r}}_i^{(g,G)}$, $i = 1,2,\cdots,N_{G}$, is compared to an individual $\bm{r}_j^{(g,P)}$ that is randomly selected from $\bm{P}^{(g)}$: if $\widehat{\bm{r}}_i^{(g,G)}$ dominates $\bm{r}_j^{(g,P)}$ as defined by (\ref{eq17}), $\bm{r}_j^{(g,P)}$ is replaced by $\widehat{\bm{r}}_i^{(g,G)}$ in the main population $\bm{P}^{(g)}$. The updating operations are continued until all the individuals in the auxiliary population $\bm{G}^{(g)}$ are compared to an individual selected from the main population $\bm{P}^{(g)}$.

\item [6)] \textbf{Termination}. The ultimate stopping criterion would be that the Pareto-front solutions of the multiple-objective routing optimization problem have been found. However, we cannot offer any proof of evidence that the Pareto-optimal routing paths have indeed been found. 

In order to have limited and predicable computational complexity, we opt for halting the optimization procedure when the pre-defined maximum affordable number of generations $g_{\max}$ has been exhausted, namely, $g=g_{\max}$, and the individuals from $\bm{A}^{(g_{\max})}$ comprise the near-Pareto solutions. Otherwise, we set $g = g + 1$ and go to 2)~\textbf{Archive}.
\end{itemize}

As a population-based nature-inspired multiple-objective optimization algorithm, the computational complexity of $\epsilon$-DMOGA is bounded by the number of generations and the population size, with some additional complexities imposed by the crossover and mutation as well as selection operations. Hence, the computational complexity can be roughly quantified by the number of cost function (CF) evaluations, which is given by $(P_{s} + N_{O})g_{\max}$ CF-evaluation.

\subsection{Convergence of $\epsilon$-DMOGA}
  As a nature-inspired multiple-objective optimization algorithm, there is randomness in the search procedures of $\epsilon$-DMOGA, hence it is quite challenge to definitely say whether a Pareto-optimal solution has been achieved. Nevertheless, $\epsilon$-DMOGA tries to ensure that the elite population archive $\bm{A}^{(g)}$ converge toward an $\epsilon$-Pareto set $\widetilde{\mathbf{R}}^{(\widetilde{\mathbf{P}})}$ in a smart distributed manner along the Pareto front. The convergence of $\epsilon$-DMOGA can be studied in a similar manner to \cite{Hanne1999onthe} by the probability of convergence, which is defined as 
\begin{align}
\lim_{g \to \infty} Pr\left(d\left(\bm{A}^{(g)},\mathbf{R}\right) \to 0 \right) = 1,
\end{align} 
where $d\left(\bm{A}^{(g)},\mathbf{R}\right)$ is a distance function between the $g$-th generation's elite population archive and the Pareto optimal $\mathbf{R}$. Additionally, the convergence of $\epsilon$-DMOGA may be also studied in a manner similar to {\it{Theorem 1: Almost sure convergence}} in \cite{Hanne1999onthe}. Hence motivated readers are referred to \cite{Hanne1999onthe} for a detailed  study of convergence in multiple-objective evolutionary algorithms.

\begin{table*}[tp!]
\vspace*{-2mm}
\caption{Distance-based adaptive coding and modulation scheme for aeronautical communications.}
\vspace*{-2mm}
\begin{center}
\resizebox{\textwidth}{!}{
\begin{tabular*}{18cm}{@{\extracolsep{\fill}}C{1.2cm}|C{2.0cm}|C{2.0cm}|C{2.0cm}|C{4.0cm}|C{4.0cm}}
\toprule
 Mode $k$ & Mode color & Modulation & Code rate & Spectral efficiency\,(bps/Hz) & Switching threshold $d_k$\,(km) \\ \toprule
 0 & None    & None   & None  & $< 0.459$ & $> 740.8$ \\ \midrule
 1 & Black   & BPSK   & 0.488 & 0.459     & 500       \\ \midrule
 2 & Magenta & QPSK   & 0.533 & 1.000     & 350       \\ \midrule
 3 & Green   & QPSK   & 0.706 & 1.322     & 200       \\ \midrule
 4 & Yellow  & 8-QAM  & 0.642 & 1.809     & 110       \\ \midrule
 5 & Blue    & 8-QAM  & 0.780 & 2.194     & 40        \\ \midrule
 6 & Cyan    & 16-QAM & 0.731 & 2.747     & 25        \\ \midrule
 7 & Red     & 16-QAM & 0.853 & 3.197     & 5.56      \\ \bottomrule
\end{tabular*}
}
\end{center}
\label{Tab1}
\vspace*{-2mm}
\end{table*}
\section{Simulation results}\label{S4}

In this section, we investigate the achievable network performance of our proposed $\epsilon$-DMOGA based multi-objective routing optimization scheme. 

\subsection{Simulated AANET}\label{S4.1}

The mobility characteristics of the nodes are critical for designing and analysing AANETs. In \cite{zhang2021semi}, we developed a semi-stochastic aircraft mobility model, which is capable of generating an arbitrary number of flights. However, ``it would be ideal to use actual node position", as stated by Kingsbury \cite{Kingsbury2009Mobile}. from Massachusetts Institute of Technology. Hence, in contrast to relying on a mobility model, which generates artificial flights and their trajectories for approximating aircraft movement, we simulate a realistic AANET in the Australian airspace based on a large-scale real historical flight data on both the busiest and quietest day of 2018. Specifically, June 29, 2018, which represents the busiest day, and December 25, 2018, which represents the quietest day. The busiest/quietest day is determined by the number of flights in air on the day, which indicates the traffic in airspace. We assume that there are $N_G=5$ GSs, namely these at Perth airport (PER), Melbourne airport (MEL), Sydney airport (SYD), Brisbane airport (BNE), and Darwin International airport (DRW). The selection of these five representative airports has jointly considered the geographical distribution and flight handling capacity. The flights considered for our investigation are real historical flights of the top-5 domestic airlines scheduled on June 29, 2018 and December 25, 2018. The top-5 domestic airlines in Australia were Quantas, Jetstar, Tigerair, Virgin Australia and Rex (Regional Express) in 2018.

The AANET employs the time division
duplexing (TDD) protocol, which has already been standardized by existing aeronautical communication systems, such as the Automatic Dependent Surveillance-Broadcast (ADS-B) \cite{Minimum2002}, L-band digital aeronautical communications
system (L-DACS) \cite{Schnell2014LDACS} and the aeronautical mobile airport
communication system (AeroMACS) \cite{Budinger2011Aeronautical}. Following the physical layer design in [31], orthogonal frequency-division multiplexing (OFDM) is adopted as the transmission technique of broadband aeronautical communications. Each aircraft has 32 transmit antennas and 4 receiver antennas. The network is allocated $B_{\text{total}}=6$\,MHz bandwidth at the carrier frequency of 5\,GHz. This bandwidth is divided into 512 subcarriers. The number of cyclic prefix (CP) samples is $N_{\text{cp}} = 32$. The transmit power per antenna is $P_t= 1$\,Watt. Furthermore, the distance-based ACM scheme of \cite{zhang2017adaptive,zhang2018regularized} designed for aeronautical communications is employed for quantifying the link quality between a pair of communicating aircraft, in which the transmit aircraft activates a specific ACM mode based on its distance from the receiver aircraft. Explicitly, our distance-based ACM scheme using $K = 8$ modes is given in Table~I, where an ACM mode is represented by a color. If the distance of two aircraft is longer than 740.8 km, there exists no adequate communication link between these two aircraft, which is marked as `None' in Table~\ref{Tab1}.  The default parameters of the AANET used in our simulations are summarised in Table~\ref{Tab2_add}.

\begin{table*}[btp!]
\vspace{-5mm}
\caption[Parameters used in this report]{Parameters used in simulating AANET}
\begin{center}
\begin{tabular}{L{4.5cm}|L{6.0cm}|L{4cm}}
\hline\hline
\multirow{13}{4.5cm}{AANET environment} & Airspace & Australian airspace \\
&\multirow{3}{6.0cm}{Airlines considered} & Top-5 domestic airlines of Quantas, Jetstar, Tigerair, \\
&& Virgin Australia and
Rex \\
&Location of GSs & PER, MEL, SYD, BNE, and DRW \\
& \multirow{2}{6.0cm}{Representative dates investigated} & December 25th, 2018 \\
& & June 29th, 2018\\
& Time period observed & 00:00 $\sim$ 24:00 \\
& Total number of flights on December 25th, 2018 & 802 \\
& Total number of flights on June 29th, 2018 & 1007\\
& Latitude &  Determined by each aircraft\\
& Longitude  &  Determined by each aircraft\\
& Altitude &  Determined by each aircraft\\
\hline
\hline
\multirow{8}{4.5cm}{Communication parameters} & Carrier frequency &  5 GHz\\
& Bandwidth  $B_{\text{total}}$ & 6 MHz \\
& Number of CPs $N_{\text{cp}}$  & 32 \\
& Number of subcarrier $N_{c\text{c}}$ & 512 \\
&  Rice factor $K_{\text{Rice}}$ & 5 dB \\
&  ACM & As detailed in Table~I \\
&  Maximum A2A communication distance & 740.8 km \\
&  Maximum A2G communication distance & 370.4 km \\
\hline
\end{tabular}
\end{center}
\label{Tab2_add}
\vspace{-5mm}
\end{table*}

\begin{table*}[tp!]
\vspace*{-1mm}
\caption{Comparing the multi-objectives of the Pareto-optimal routing paths for flight TT589 on June 29, 2018.}
\vspace*{-1mm}
\begin{center}
\resizebox{\textwidth}{!}{
\begin{tabular*}{16.0cm}{@{\extracolsep{\fill}}C{3.0cm}|C{2.5cm}C{2.5cm}|C{2.5cm}C{2.5cm}}
\toprule
 & 2-hop solution 1 & 2-hop solution 2 & 3-hop solution 1 & 3-hop solution 2 \\ \midrule
 $\mathcal{J}_1$ SE\,(bps/Hz) & 0.459      & 0.459      & 1.000      & 1.000 \\
 $\mathcal{J}_2$ delay\,(s)   & 0.01500361 & 0.01500365 & 0.03000395 & 0.03000379 \\
 $\mathcal{J}_3$ PET\,(s)     & 1902.86767 & 2586.56434 & 1900.42115 & 1132.96596
\\ \bottomrule
\end{tabular*}
}
\end{center}
\label{Tab2}
\vspace*{-1mm}
\end{table*}

\begin{figure*}[htbp]
\vspace{-4mm}
\begin{center}
\vspace{-2mm}
 \subfigure[No 1 hop to a GS]{
  \includegraphics[width=0.45\textwidth]{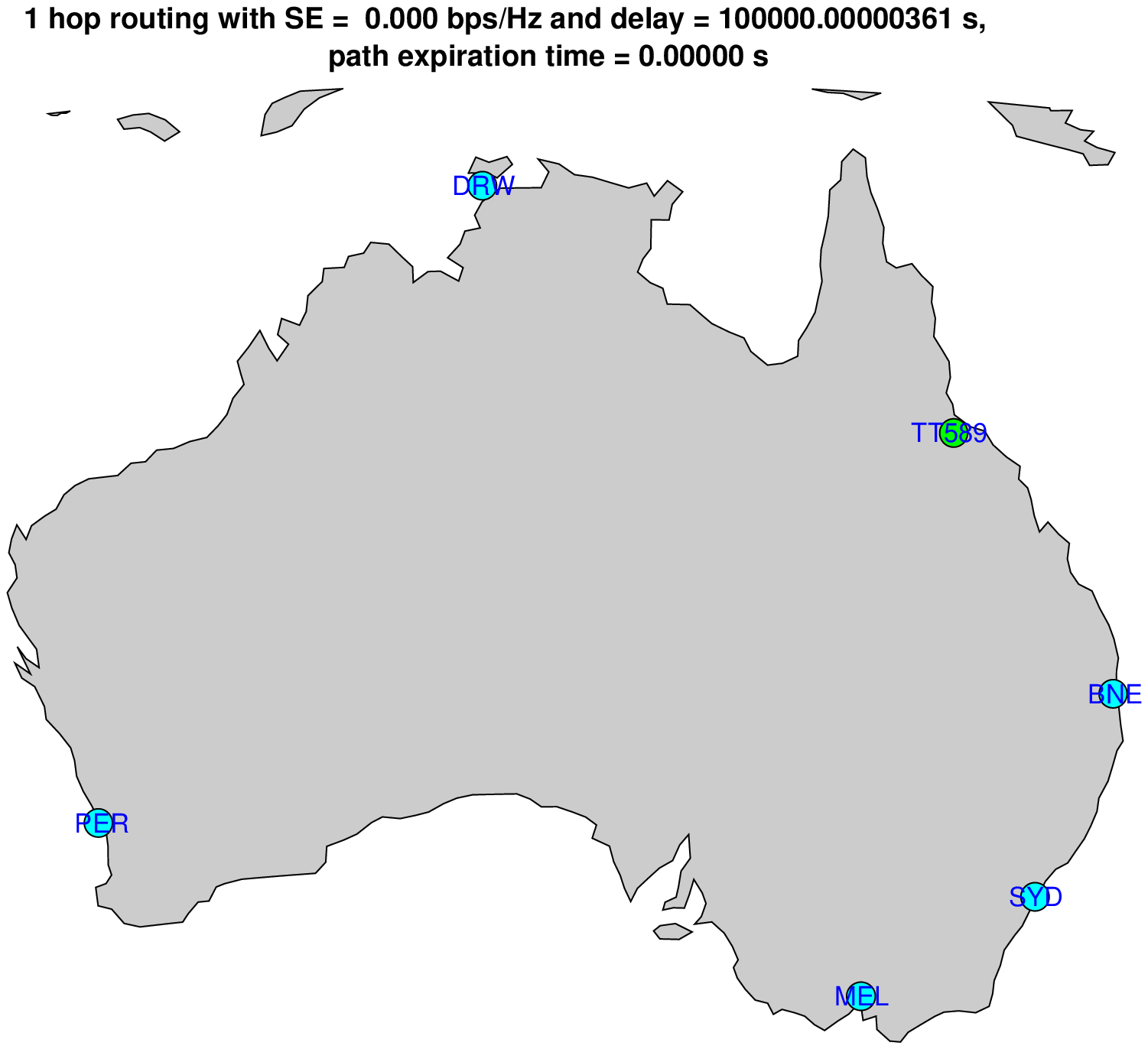} 
  \label{fig3a}
 }%
 \\ 
 \vspace{-2mm}
 \subfigure[2 hops to a GS -- Solution-1]{
  \includegraphics[width=0.45\textwidth]{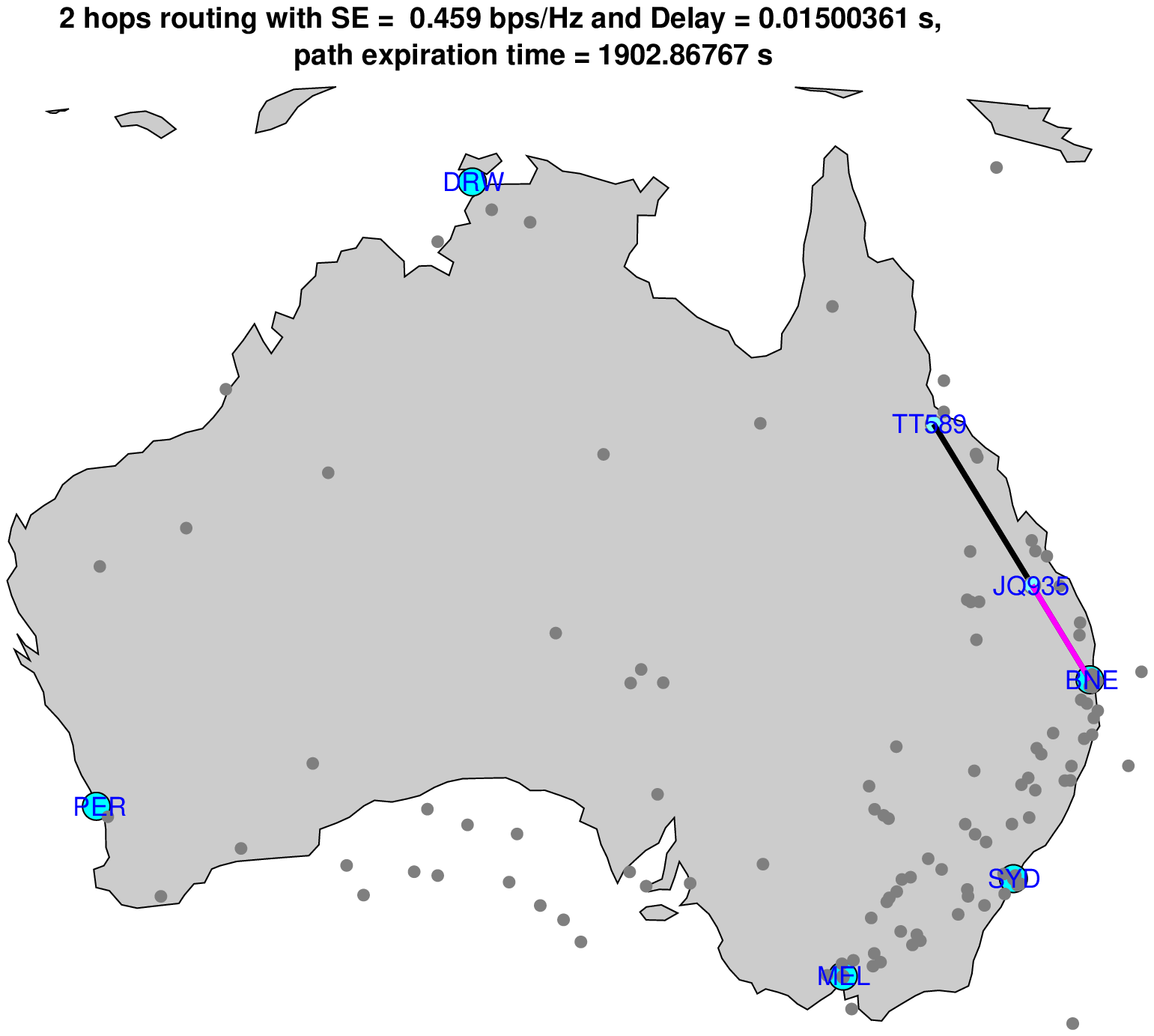}
  \label{fig3b}
 }\hspace*{5mm}
 \vspace{-2mm}
 \subfigure[2 hops to a GS -- Solution-2]{
  \includegraphics[width=0.45\textwidth]{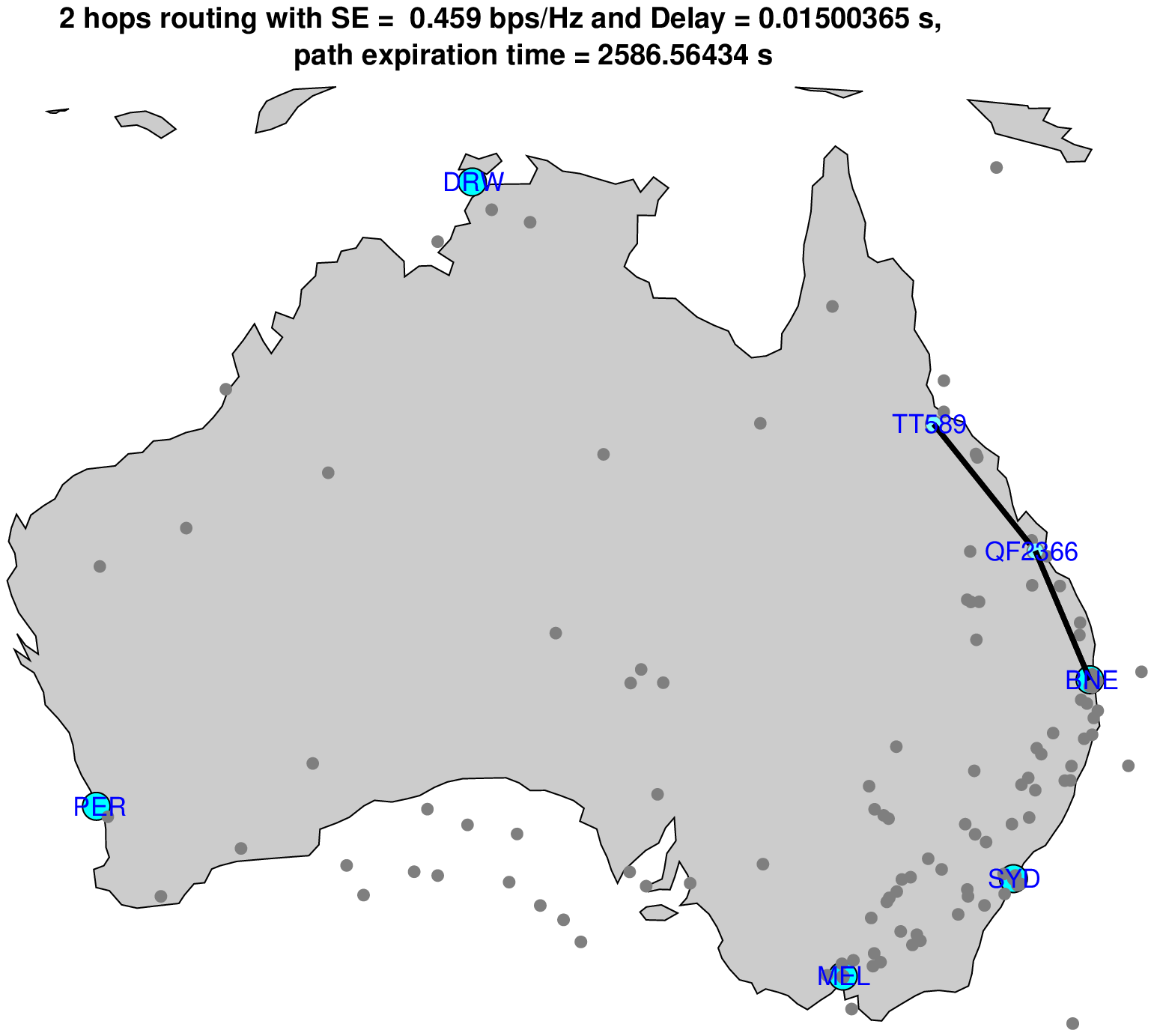} 
  \label{fig3c}
 }%
 \\
 \vspace{-1mm}
 \subfigure[3 hops to a GS -- Solution-1]{
  \includegraphics[width=0.45\textwidth]{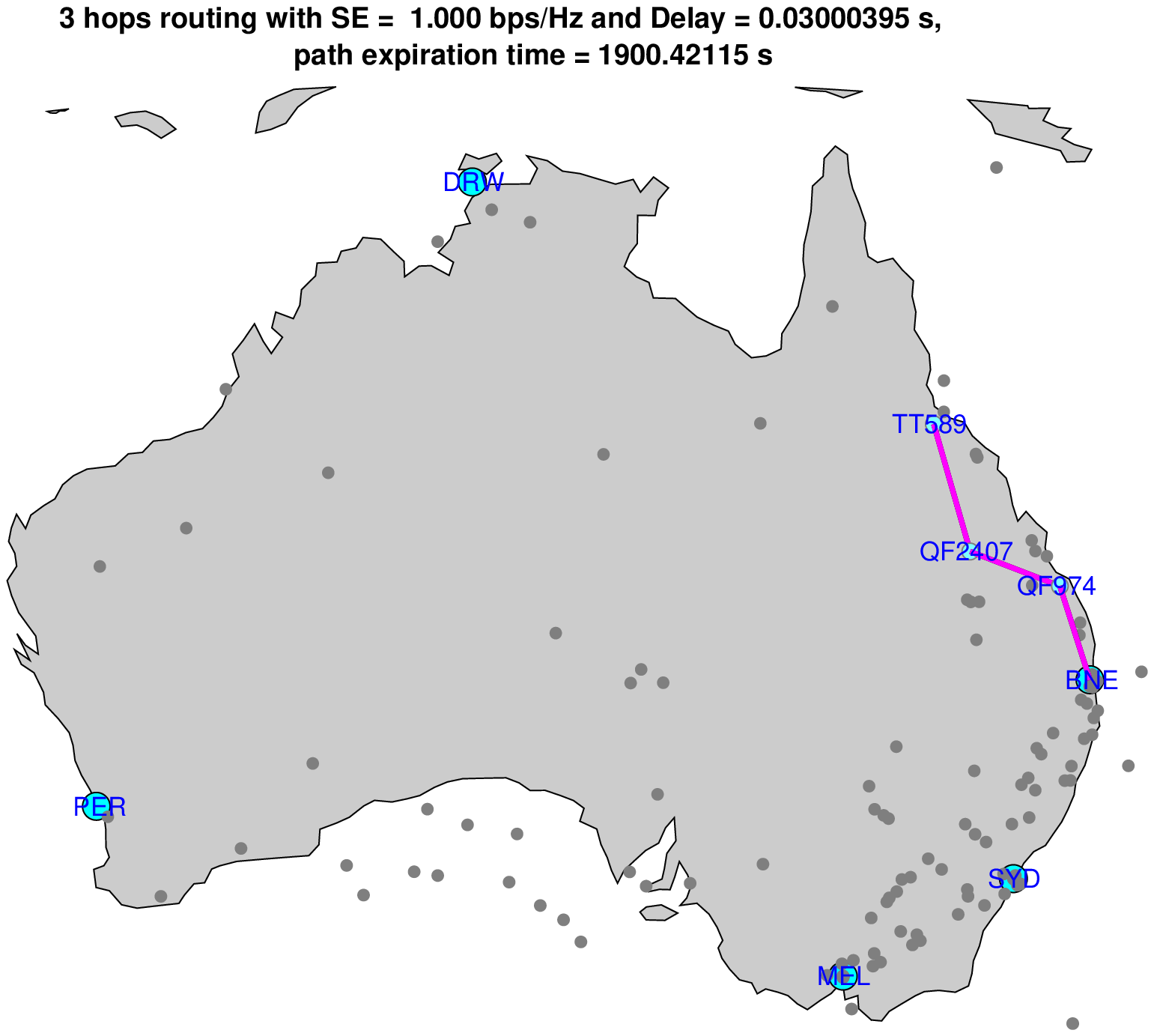}
  \label{fig3d}
 }\hspace*{5mm}
 \vspace{-1mm}
  \subfigure[3 hops to a GS --5 Solution-2]{
  \includegraphics[width=0.46\textwidth]{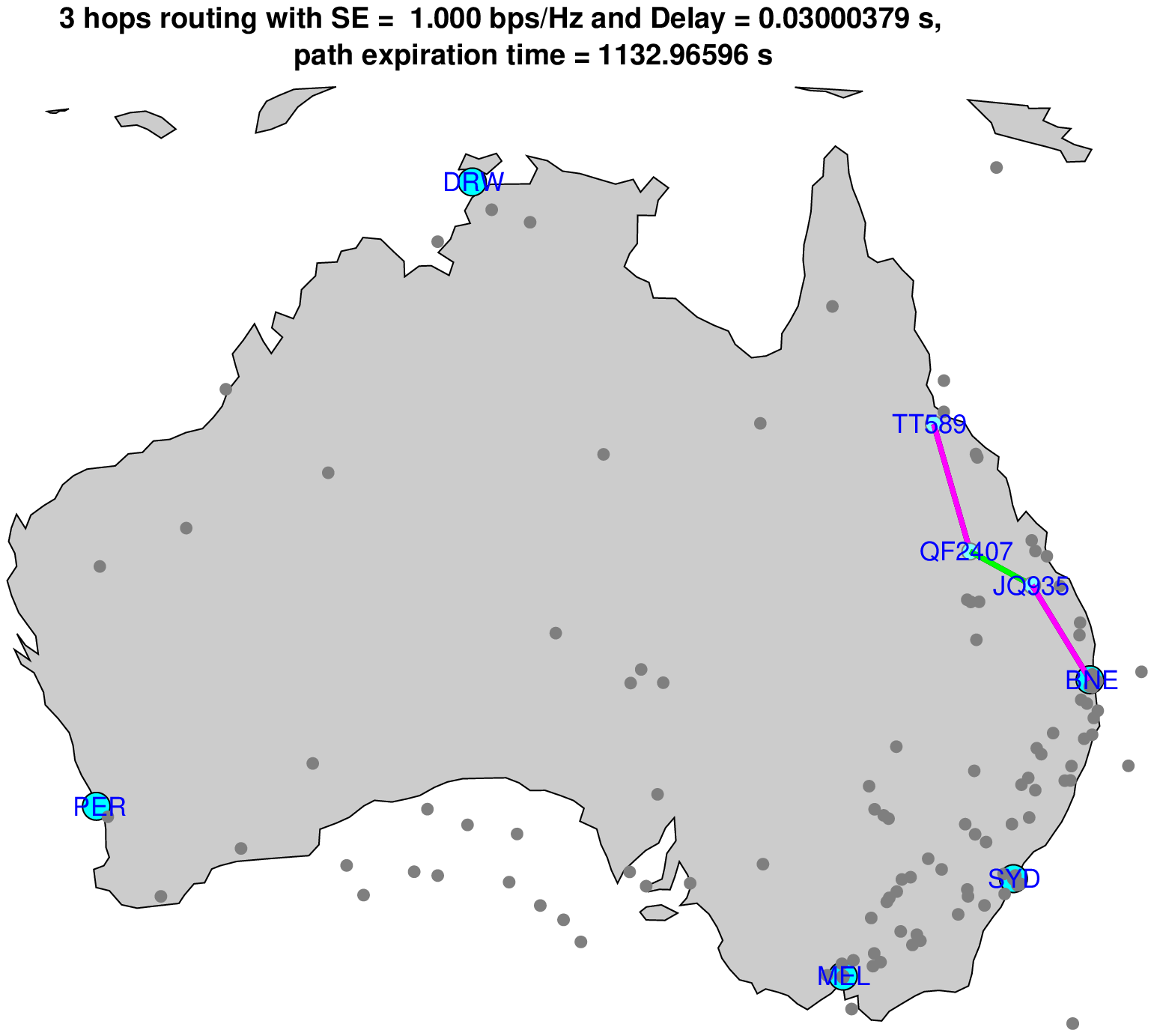}
  \label{fig3e}
 }
\end{center}
\vspace{-4mm}
\caption{A specific example of the flight-TT589's routing path to a GS, all of which are Pareto-optimal compared to any of other routing paths.}
\label{fig3}
\vspace{-4mm}
\end{figure*}

\subsection{A specific example of the flight TT589}\label{S4.2}

First, we investigate the achievable multiple objectives of the network layer performance via the specific example of the flight TT589 on June 29, 2018, which may be extrapolated to other flights and other dates. As shown in Fig.~\ref{fig3a}, there is no available direct link to any GS  at the five airports considered. However, if we increase the number of the affordable relay nodes and tolerate a higher delay, there are available routing paths to the GS at Brisbane airport. As shown in Fig.~\ref{fig3b} and Fig.~\ref{fig3c}, there are two Pareto-optimal 2-hop routing paths to the GS deployed at Brisbane airport relying on the relay nodes JQ935 and QF2366, respectively. Both routing paths have a spectrum efficiency (SE) of 0.459\,bps/Hz according to the SE of the `black' link of Table~\ref{Tab1}. The multi-objective functions of these two solutions are compared in the left half of Table~\ref{Tab2}. It is clear that the two routing paths shown in Fig.~\ref{fig3b} and Fig.~\ref{fig3c} do not dominate each other, and hence both are Pareto-optimal routing paths, provided that the affordable number of relay nodes is one. 

If the affordable number of relay nodes is two, we can find routing paths having a higher SE. As shown in Fig.~\ref{fig3d} and Fig.~\ref{fig3e}, there are two Pareto-optimal routing paths to the GS at Brisbane airport via three hops, namely, the routing path TT589$\rightarrow$QF2407$\rightarrow$QF974$\rightarrow$BNE and the routing path TT589$\rightarrow$QF2407$\rightarrow$QF935$\rightarrow$BNE, respectively. Both the routing paths have a SE of 1.000\,bps/Hz according to the SE of the `magenta' link, which determines the end-to-end SE. As confirmed in the right half of Table~\ref{Tab2}, these two 3-hop routing paths are also Pareto-optimal, as they do not dominate each other.

\begin{figure*}[tp!]
\vspace{-2mm}
\begin{center}
 \subfigure[June 29, 2018]{
  \includegraphics[width=1.0\columnwidth]{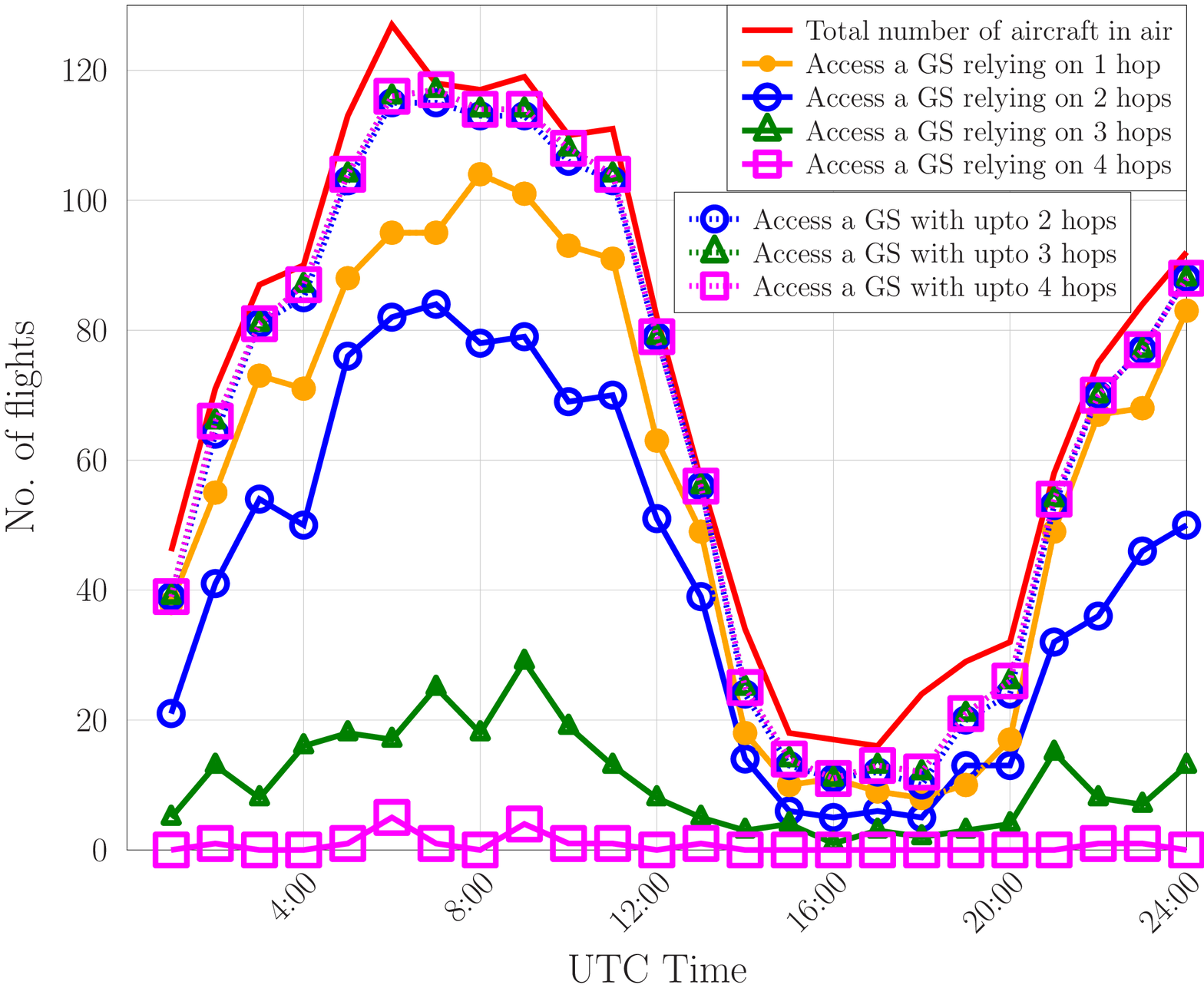} 
  \label{fig4a}
 }%
 \subfigure[December 25, 2018]{
  \includegraphics[width=1.0\columnwidth]{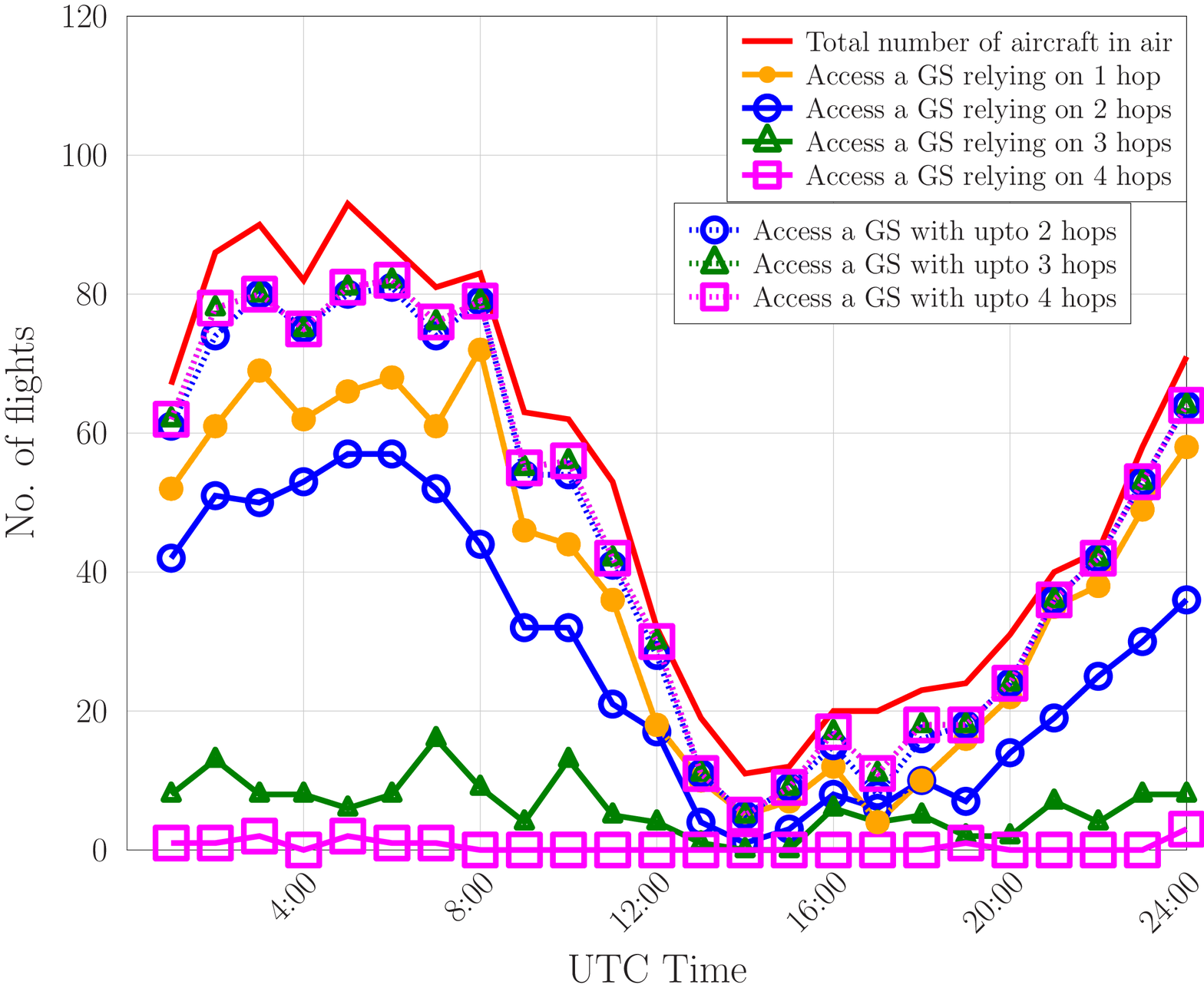}
  \label{fig4b}
 }
\end{center}
\vspace{-2mm}
\caption{The number of flights in air over 24 hours capable of accessing the Internet using the $\epsilon$-DMOGA multi-objective routing optimization.}
\label{fig4}
\vspace{-2mm}
\end{figure*}

\subsection{Overall multi-objective network-layer performance}\label{S4.3}

We now investigate the overall network-layer performance, including how many flights in the air over a period of 24 hours can access the Internet through our $\epsilon$-DMOGA based multi-objective routing as well as the average end-to-end SE, the average end-to-end latency and the routing paths' average PET, on June 29, 2018 and December 25, 2018, respectively.

\subsubsection{The number of flights that can access the Internet}

Clearly, the number of flights in air changes over the 24 hours of the day, as shown in Fig.~\ref{fig4a} and Fig.~\ref{fig4b} for June 29, 2018 and December 25, 2018, respectively. The peak number of flights in air occurs at UTC time 06:00, namely 127 flights, whilst the lowest number of the flights in air occurs at UTC time 17:00, namely 16 flights. We start by investigating the number of flights that can access a GS at any one of the five airports considered. Explicitly, in Fig.~\ref{fig4}, we investigate how many flights can access a GS relying on 1 hop, 2 hops, 3 hops and 4 hops, respectively, using our $\epsilon$-DMOGA based multi-objective routing optimization scheme. Additionally, we also provides the numbers of flights that can access a GS with up to 2 hops, up to 3 hops and up to 4 hops, respectively, obtained by the $\epsilon$-DMOGA based multi-objective routing optimization. 

As shown by the solid lines of Fig.~\ref{fig4a} and Fig.~\ref{fig4b}, most of the flights in air can access a GS relying on 1 hop, i.e. via a direct link. Furthermore, by analyzing the results obtained using the $\epsilon$-DMOGA scheme, it can be found that there is a significant number of flights which can achieve higher end-to-end SE than the single-hop paths by relying on 2 hops.  By contrast, seldom flights can achieve higher end-to-end SE relying on 3 hops and 4 hops than those relying on 2 hops. Although it is not explicitly shown by the figures, we found that almost all flights can access a GS with up to 2 hops. Hence, increasing the affordable number of hops to 3 and 4 contributes little to the total number of flights that can access a GS, regardless of the hour of the day.

\begin{table*}[tbp!]
\vspace*{-1mm}
\caption{Comparing end-to-end spectrum efficiency of one-hop Pareto-optimal routing solutions and Pareto-optimal routing solutions with up to two hops at UTC time 18:00 on June 29, 2018.}
\vspace*{-2mm}
\begin{center}
\resizebox{\textwidth}{!}{
\begin{tabular*}{17.0cm}{@{\extracolsep{\fill}}c|rrrrrrrr|c}
\toprule
 & \multicolumn{8}{c|}{Individual flights' SE\,(bps/Hz)} & Average SE\,(bps/Hz) \\ \midrule
8 flights with 1 hop & 1.8090 & 1.3220 & 2.1940 & 2.1940 & 1.8090 & 1.0000 & 1.0000 & 1.3220 & 1.58125 \\ \midrule
10 flights with up   & 1.8090 & 1.3220 & 2.1940 & 2.1940 & 2.1940 & 1.3220 & 1.3220 & 1.3220 & 1.52282 \\
to 2 hops            & 0.7295 & 0.8197 &        &        &        &        &        &        & \\ \bottomrule
\end{tabular*}
}
\end{center}
\label{Tab3}
\vspace*{-1mm}
\end{table*}

\begin{figure*}[tp!]
\vspace{-2mm}
\begin{center}
 \subfigure[June 29, 2018]{
  \includegraphics[width=1.0\columnwidth]{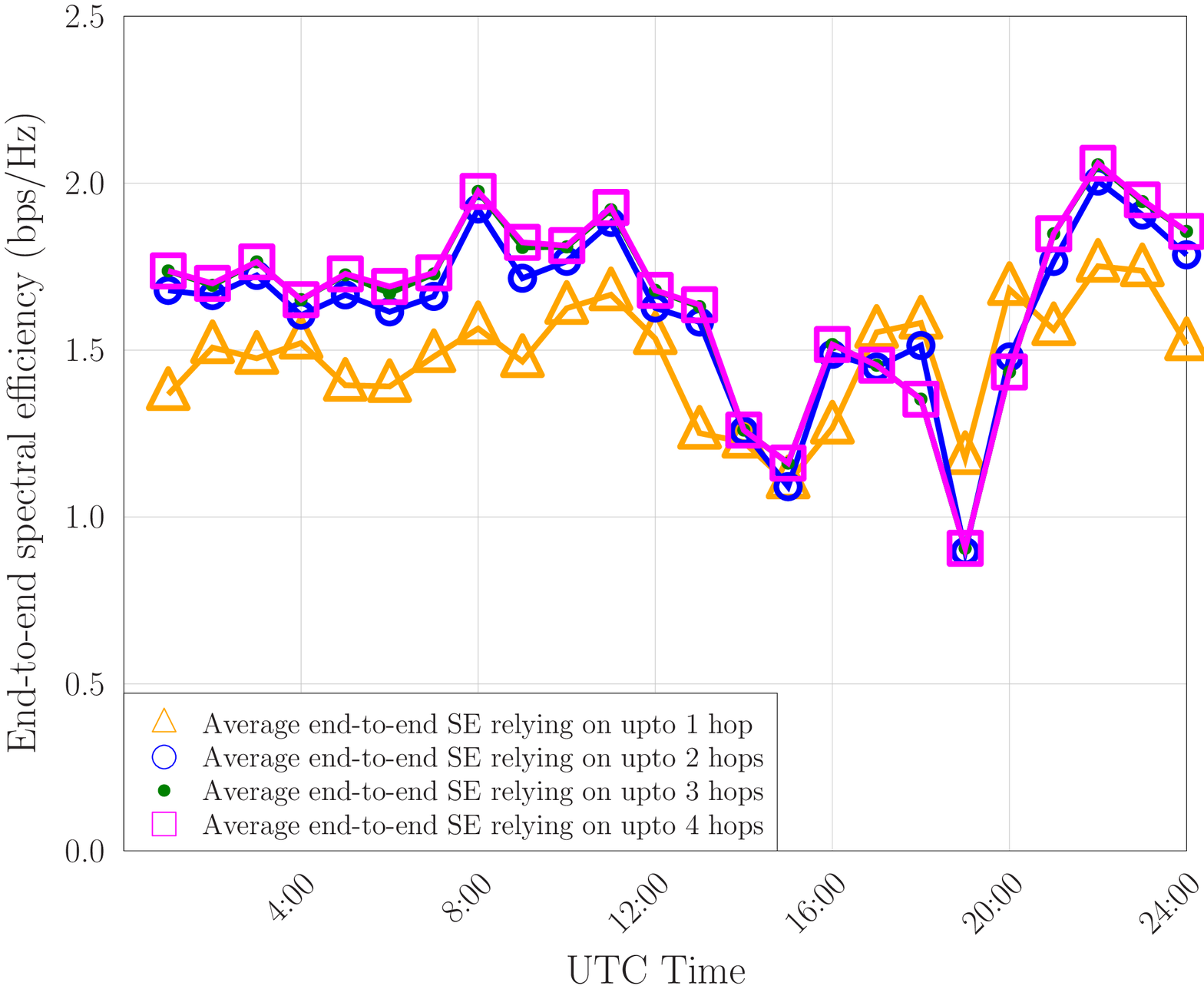} 
  \label{fig5a}
 }%
 \subfigure[December 25, 2018]{
  \includegraphics[width=1.0\columnwidth]{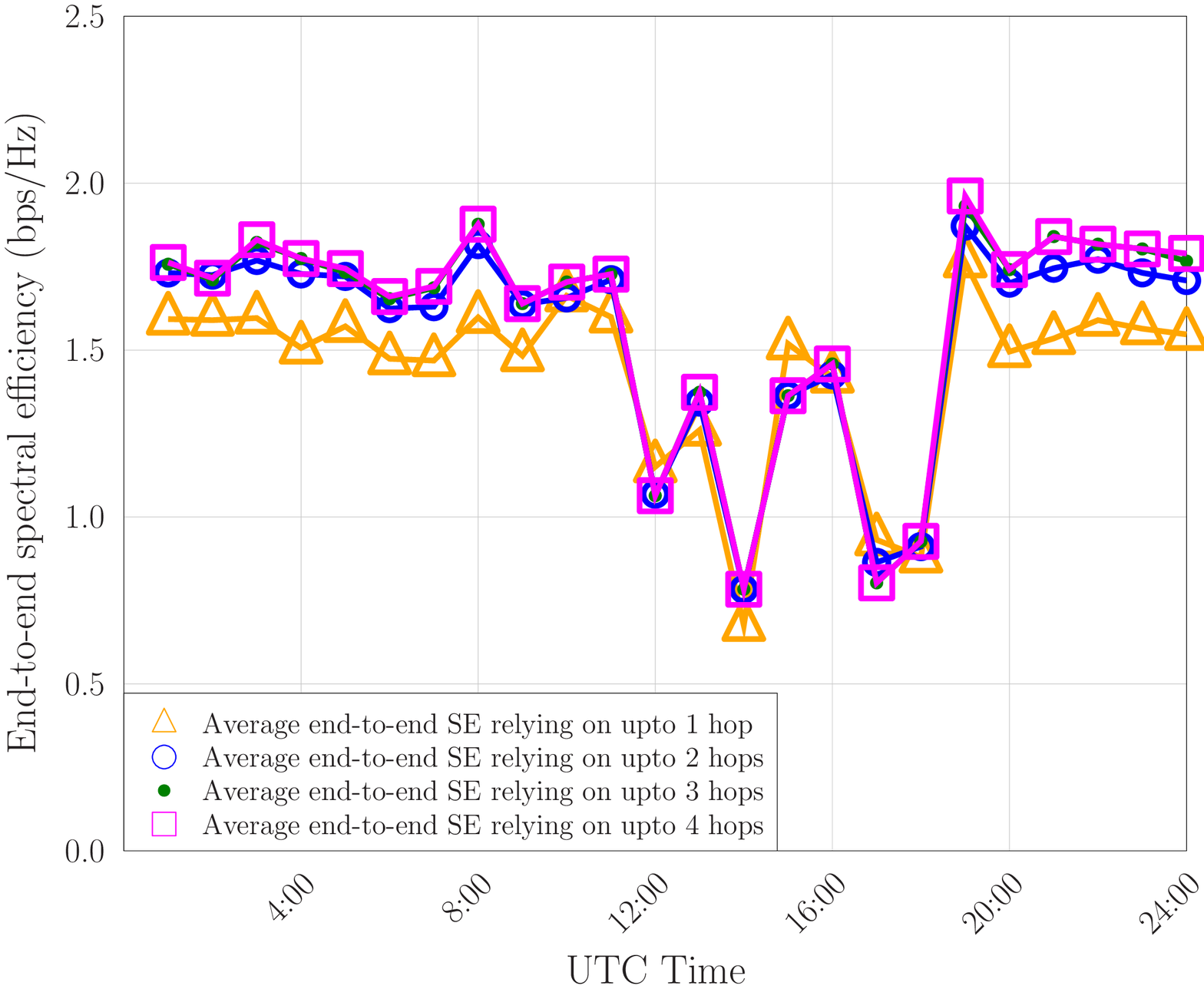}
  \label{fig5b}
 }
\end{center}
\vspace{-2mm}
\caption{The average end-to-end spectral efficiency achieved by the $\epsilon$-DMOGA based multi-objective routing optimization.}
\label{fig5}
\vspace{-2mm}
\end{figure*}

\subsubsection{Average end-to-end SE}

Since the routes of the same number of hops for different flights have different end-to-end SEs, Fig.~\ref{fig5a} and Fig.~\ref{fig5b} portray the achievable average end-to-end SE over 24 hours on June 29, 2018 and December 25, 2018, respectively. As shown in Fig.~\ref{fig5a} and Fig.~\ref{fig5b}, the routing paths relying on up to 2, 3 and 4 hops are capable of achieving somewhat higher average end-to-end SE than those relying on 1 hop most times of the day except for UTC times 17:00, 18:00, 19:00 and 20:00 on June 29 as well as UTC time at 15:00 on December 25, mainly in the scenarios of having low flight times. But naturally, this SE improvement is achieved at the cost of a higher delay. To elaborate a little further, it is unexpected to encounter a lower SE for 2 hops than for 1 hop, because a single hop tends to be longer, which results in a lower SE. The issue here is that there are more Pareto-optimal routing paths of up to 2 hops than the number of Pareto-optimal 1-hop routes. We illustrate this point using the example at UTC time 18:00 on June 29, 2018. There are 8 flights having direct links to a GS,
while there are 10 flights having routing paths to a GS relying on up to 2 hops -- some relying on direct link and some relying on 2 hops to access a GS. Table~\ref{Tab3} compares the individual flights' SEs of these two groups as well as their average SEs. Although the average SE of the second group is lower than that of the first group, observe that eight of the ten Pareto-optimal routing paths with up to 2 hops have the same or higher SEs than those of the eight 1-hop routing paths.

Extrapolating from the results of the busiest day and the quietest day depicted in Fig.~\ref{fig5a} and Fig.~\ref{fig5b}, we may draw the conclusion that the AANET over the Australian airspace can achieve an average SE of about 1.2\,bps/Hz for low-flight-times during the period of UTC 12:00-20:00, while its average SE increases to around 1.7\,bps/Hz for high-flight-times.

\begin{figure*}[tbp!]
\vspace{-2mm}
\begin{center}
 \subfigure[June 29, 2018]{
  \includegraphics[width=1\columnwidth]{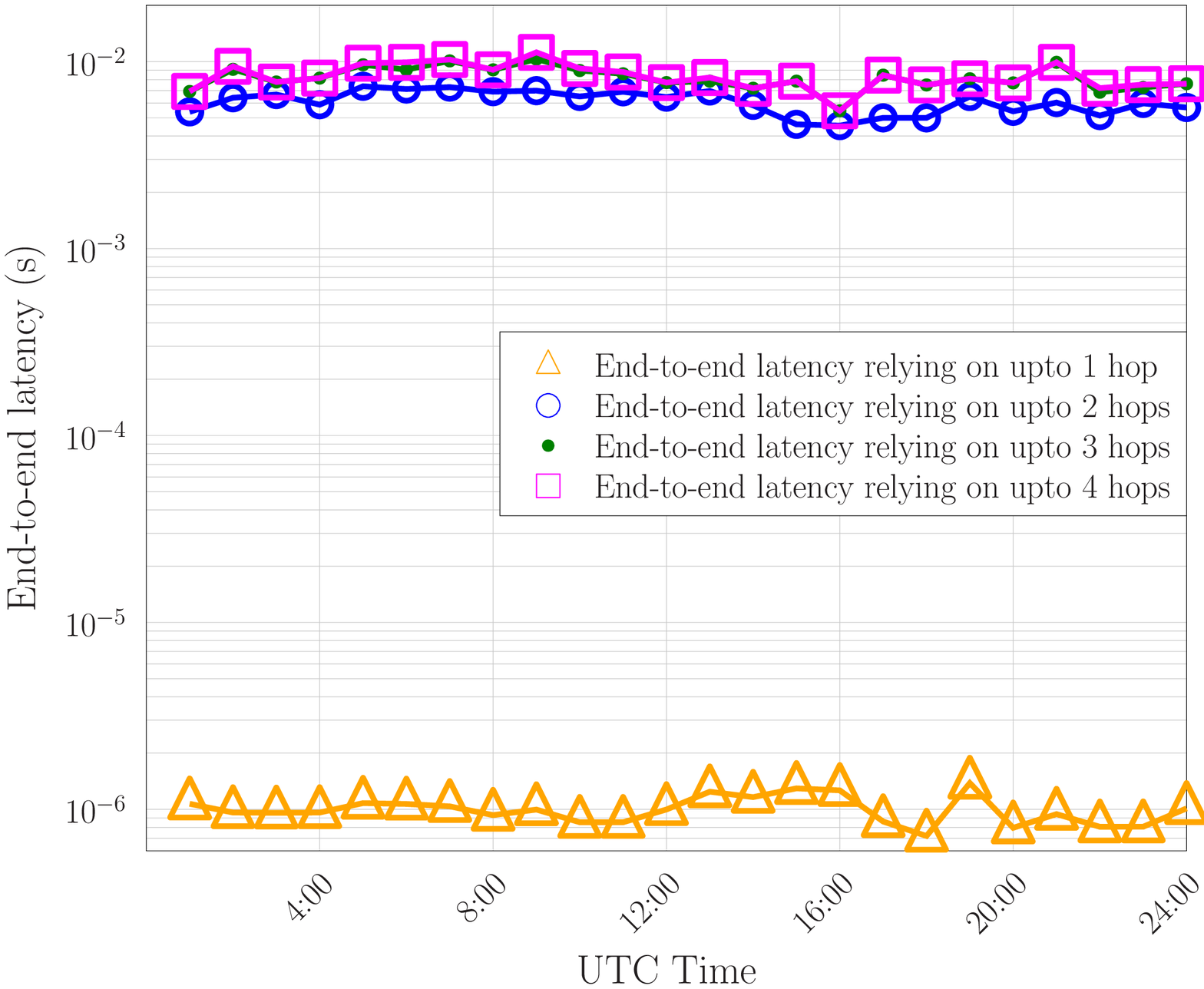} 
  \label{fig6a}
 }%
 \subfigure[December 25, 2018]{
  \includegraphics[width=1\columnwidth]{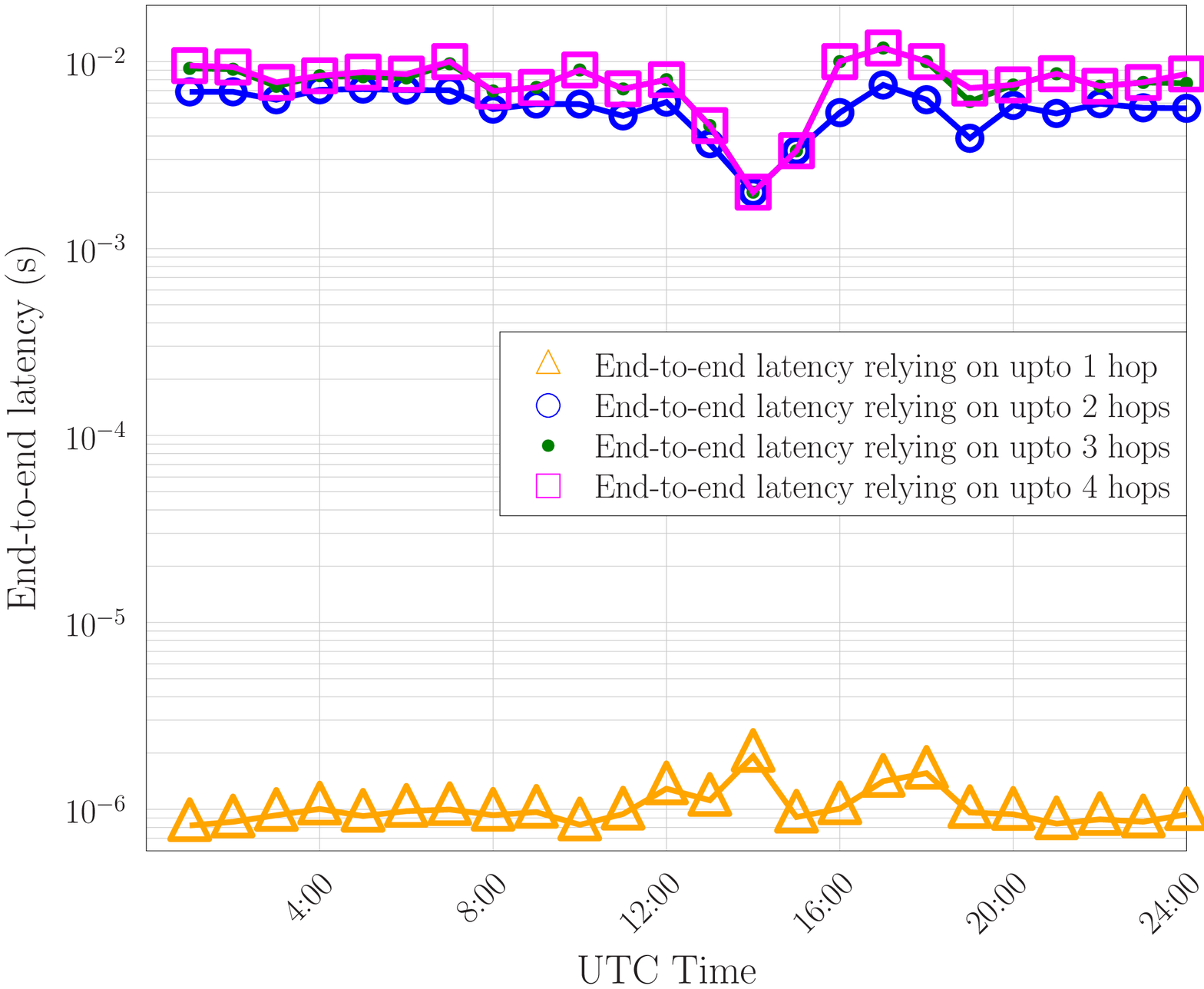}
  \label{fig6b}
 }
\end{center}
\vspace{-2mm}
\caption{The average end-to-end latency achieved by the $\epsilon$-DMOGA based multi-objective routing optimization.}
\label{fig6}
\vspace{-2mm}
\end{figure*}

\subsubsection{Average end-to-end latency} 

The average end-to-end latency imposed by the Pareto-optimal routing paths over 24 hours on June 29, 2018 and December 25, 2018 are depicted in Fig.~\ref{fig6a} and Fig.~\ref{fig6b}, respectively. As expected, the direct links have the lowest latency, since they only have single-hop propagation delay upon accessing a GS. As seen in Fig.~\ref{fig6a} and Fig.~\ref{fig6b}, the average end-to-end latency relying on up to 2 hops is significantly higher than that relying on direct links. This is because the end-to-end latency of a 2-hop routing path also includes the signal processing delay and queuing delay, which are significantly higher than the propagation delay. Furthermore, the average end-to-end latency relying on up to 3 hops and up to 4 hops is  higher than that relying on up to 2 hops. But the difference between the average end-to-end latency relying on up to 3 hops and up to 4 hops is small, because the number of routing paths relying on 4 hops is relatively small, as shown in Fig.~\ref{fig4a} and Fig.~\ref{fig4b}.

As observed from Fig.~\ref{fig6a} and Fig.~\ref{fig6b}, the variation of the average end-to-end latency over 24 hours of a day is small. Furthermore, by extrapolating from the results of Fig.~\ref{fig6a} and Fig.~\ref{fig6b}, we may conclude that the average end-to-end latency may be as low as 0.01\,s in the AANET over the Australian airspace, provided that a link is available.

\begin{figure*}[tp!]
\vspace{-2mm}
\begin{center}
 \subfigure[June 29, 2018]{
  \includegraphics[width=1\columnwidth]{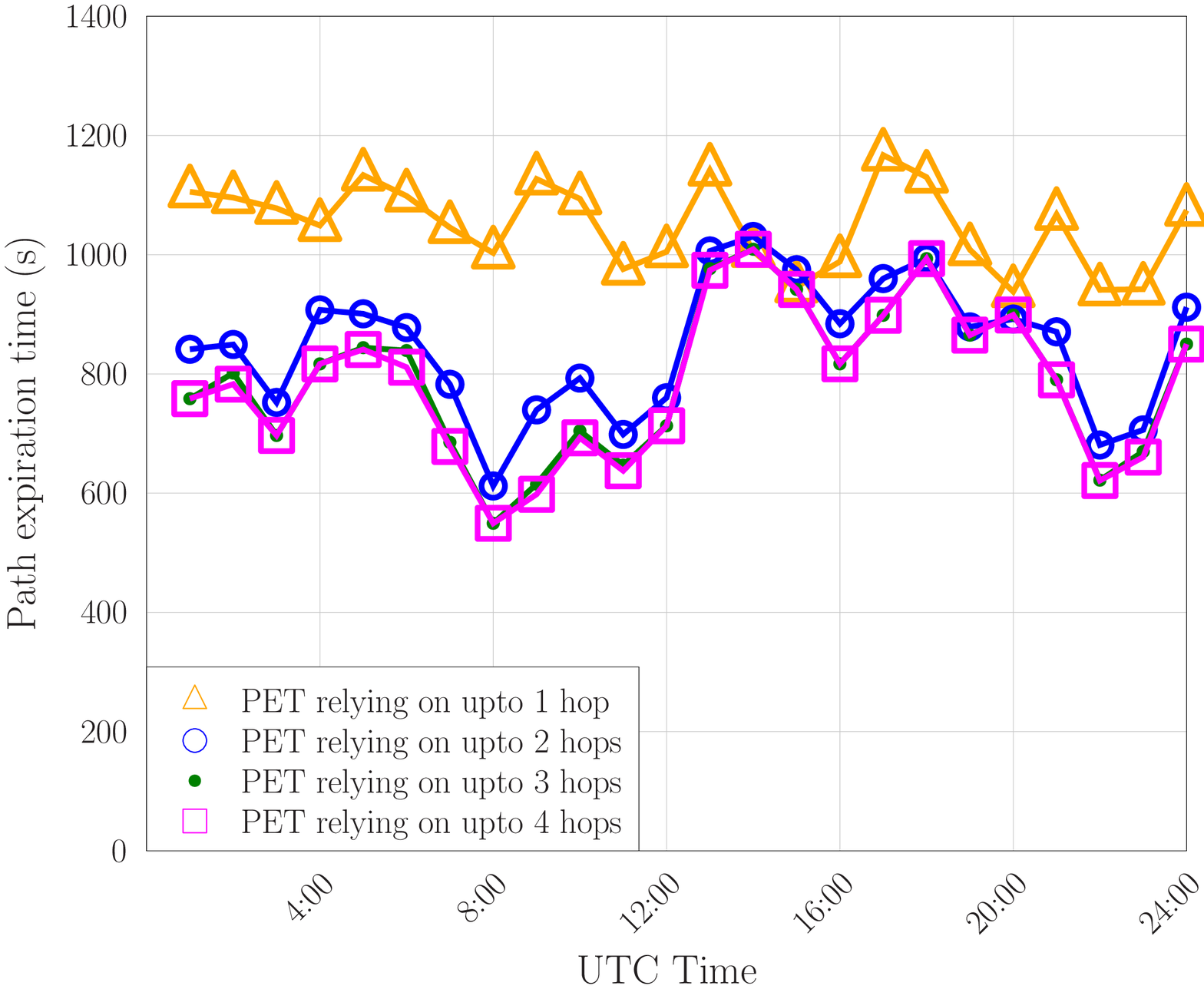} 
  \label{fig7a}
 }%
 \subfigure[December 25, 2018]{
  \includegraphics[width=1\columnwidth]{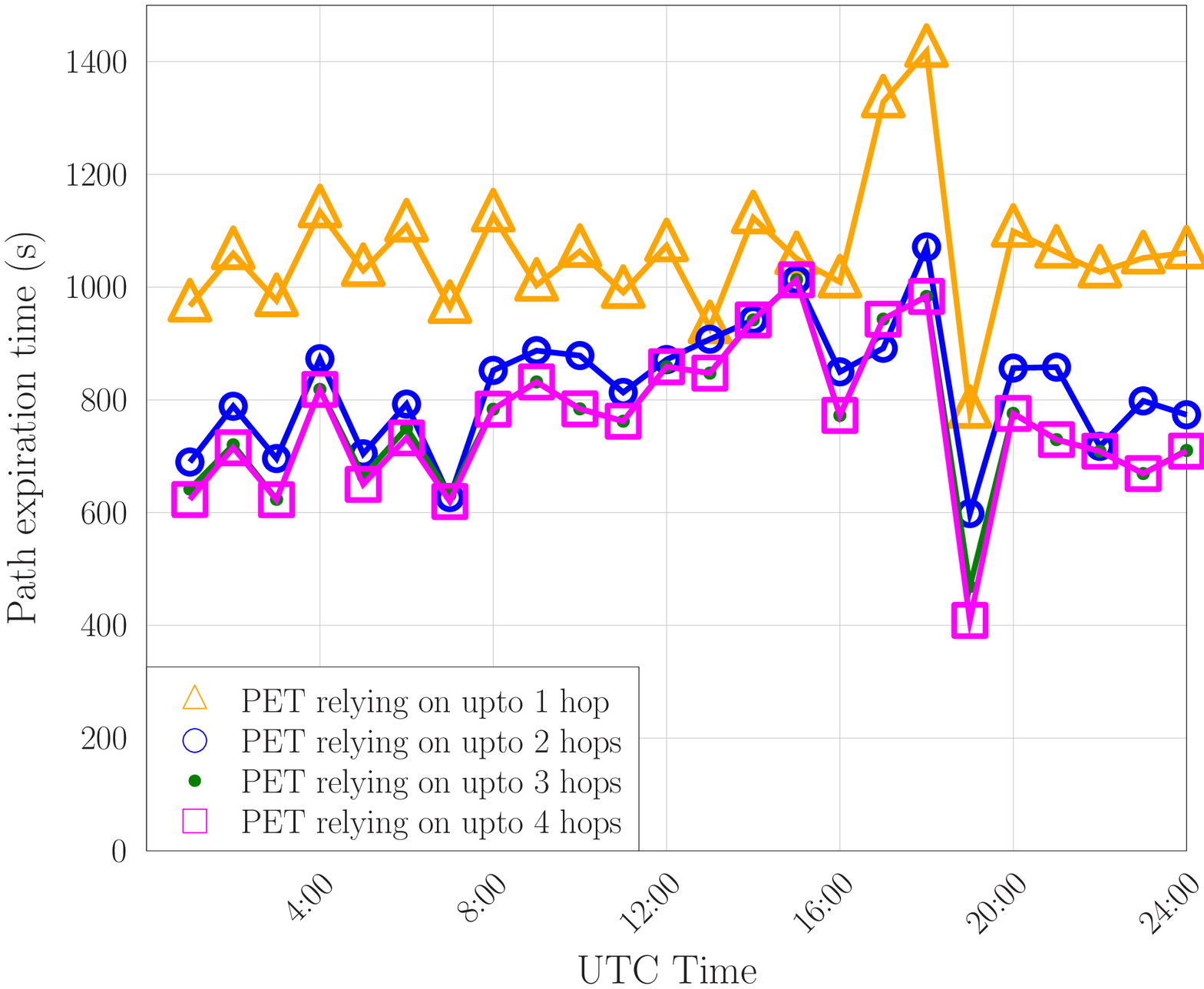}
  \label{fig7b}
 }
\end{center}
\vspace{-2mm}
\caption{The average path expiration time achieved by the $\epsilon$-DMOGA based multi-objective routing optimization.}
\label{fig7}
\vspace{-2mm}
\end{figure*}

\subsubsection{Average path expiration time}

The average PETs over 24 hours on June 29, 2018 and December 25, 2018 are portrayed in Fig.~\ref{fig7a} and Fig.~\ref{fig7b}, respectively. Intuitively, the routing path may become vulnerable to potential breakage upon increasing the number of hops. Our investigations based on real historical flight data both on June 29, 2018 and on December 25, 2018 confirm  this intuition. Explicitly, as shown in Fig.~\ref{fig7a} and Fig.~\ref{fig7b}, the routing paths relying on a direct link to a GS have considerably longer average PET than those relying on 2 , 3  and 4 hops, except around UTC time 14:00 on June 29, 2018. Furthermore, the routing paths relying on 2 hops have longer average PET with a high probability than those relying on 3  and 4 hops, but the difference between the average PETs relying on 3 and 4 hops are hardly noticeable, since the number of routing paths relying on 4 hops is very low.

It can be seen that the average PET varies over the 24 hours of a day. On average, we may extrapolate that the PET is around 800\,s i.e. over minutes in the AANET over the Australian airspace.

\begin{table*}[tp!]
\vspace*{-2mm}
\caption{the achievable network performance of the AANET over Australia airspace}
\vspace*{-2mm}
\begin{center}
\begin{tabular}{C{4.0cm}C{4.0cm}C{4.0cm}}
\toprule
 Average spectrum efficiency & Average end-to-end latency & Average  path expiration time \\ \midrule
 {1.2\,bps/Hz at low-flight times} {1.7\,bps/Hz at high-flight times} & 0.01\,s & 800\,s $\approx$ 13.3 min \\ \bottomrule
\end{tabular}
\end{center}
\label{Tab4}
\vspace*{-2mm}
\end{table*}

\subsubsection{Summary}

Based on the results obtained using real historical flight data on two representative dates in 2018, namely on the busiest day of June 29 and on the quietest day of December 25, we may extrapolate the achievable network layer performance for the AANET over the Australian airspace using our $\epsilon$-DMOGA based multi-objective routing optimization. The overall network performance is summarized in Table~\ref{Tab4} at a glance.

\section{Conclusions}\label{S5}

In order to provide Internet service above the clouds, an $\epsilon$-DMOGA based multiple-objective routing optimization has been developed by taking into account the unique features of routing problem in AANETs. Explicitly, the end-to-end SE, the end-to-end latency and the PET have been jointly optimized by the proposed $\epsilon$-DMOGA for determining the Pareto-optimal routing paths. The achievable end-to-end SE, end-to-end latency and PET performance using our $\epsilon$-DMOGA based multiple-objective routing optimization have been investigated based on the top-5 Australia domestic airlines' real historical flight data on two representative dates in  2018, namely the busiest day of June 29 and the quietest day of December 25, in term of the number of flights in air. Our simulation results have quantified the networking capability of the AANET over the Australian airspace. Furthermore, our investigations have also offered useful design considerations for the AANETs in other parts of the worlds.

%\linespread{1.30}
\bibliographystyle{IEEEtran}

\end{document}